\documentclass[11pt]{article}

\usepackage[letterpaper,margin=1in]{geometry}
\usepackage[T1]{fontenc}
\usepackage{mathptmx}
\usepackage{microtype}
\usepackage{amsmath,amssymb,amsthm,mathtools}
\usepackage{booktabs}
\usepackage{tikz}
\usepackage{enumitem}
\usepackage{makecell}
\usepackage{algorithm}
\usepackage[noend]{algpseudocode}
\algrenewcommand{\algorithmicrequire}{\textbf{Input:}}
\algrenewcommand{\algorithmicensure}{\textbf{Output:}}
\algnewcommand{\FixedData}{\item[\textbf{Fixed data:}]}
\usepackage{xcolor}
\usepackage{hyperref}
\usepackage[nameinlink,capitalize,noabbrev]{cleveref}
\crefname{algorithm}{Algorithm}{Algorithms}
\Crefname{algorithm}{Algorithm}{Algorithms}

\DeclareMathAlphabet{\mathcal}{OMS}{cmsy}{m}{n}

\hypersetup{
  colorlinks=true,
  linkcolor=blue!55!black,
  citecolor=blue!55!black,
  urlcolor=blue!55!black,
  pdftitle={Optimal Separations Between Randomized and Quantum Query Complexity under Permutation Symmetry},
  pdfauthor={Yunqi Huang, Lvzhou Li, Zekun Ye, Shenggen Zheng}
}

\setlist{nosep,leftmargin=*}
\allowdisplaybreaks

\newtheorem{theorem}{Theorem}[section]
\newtheorem{lemma}[theorem]{Lemma}
\newtheorem{proposition}[theorem]{Proposition}
\newtheorem{corollary}[theorem]{Corollary}
\newtheorem*{stringrestatement}{Theorem~\ref{thm:string-main} (restated)}
\newtheorem*{hypergraphrestatement}{Theorem~\ref{thm:hypergraph-main} (restated)}
\newtheorem*{lowerrestatement}{Theorem~\ref{thm:lower-main} (restated)}
\theoremstyle{definition}
\newtheorem{definition}[theorem]{Definition}
\newtheorem{remark}[theorem]{Remark}
\newtheorem{problem}{Problem}
\crefname{problem}{Problem}{Problems}
\Crefname{problem}{Problem}{Problems}

\newcommand{\cA}{\mathcal{A}}

\newcommand{\cD}{\mathcal{D}}

\newcommand{\cH}{\mathcal{H}}
\newcommand{\cL}{\mathcal{L}}
\newcommand{\cP}{\mathcal{P}}
\newcommand{\cR}{\mathcal{R}}

\newcommand{\E}{\mathbb{E}}

\newcommand{\im}{\operatorname{im}}
\newcommand{\dom}{\operatorname{dom}}

\newcommand{\eps}{\varepsilon}
\newcommand{\ket}[1]{\lvert #1\rangle}
\newcommand{\bra}[1]{\langle #1\rvert}

\newcommand{\norm}[1]{\left\lVert #1\right\rVert}
\newcommand{\abs}[1]{\left\lvert #1\right\rvert}

\newcommand{\one}{\mathbf{1}}
\newcommand{\cO}{\mathcal O}

\begin{document}

\hypersetup{pageanchor=false}
\pagenumbering{roman}
\begin{titlepage}
  \thispagestyle{empty}
  \centering
  \vspace*{0.15in}
  {\LARGE\bfseries Optimal Separations Between Randomized and Quantum Query Complexity under Permutation Symmetry \par }
  \vspace{0.3in}
  \renewcommand{\thefootnote}{\fnsymbol{footnote}}
  {\large Yunqi Huang\footnotemark[1] \quad Lvzhou Li\footnotemark[2] \quad Zekun Ye\footnotemark[3] \quad Shenggen Zheng\footnotemark[4]\par}
  \footnotetext[1]{\raggedright Quantum Science Center of Guangdong-Hong Kong-Macao Greater Bay Area, Shenzhen, China; College of Physics and Optoelectronic Engineering, Shenzhen University, Shenzhen, China. Email: \href{mailto:huangyunqi29@gmail.com}{\texttt{huangyunqi29@gmail.com}}.}
  \footnotetext[2]{\raggedright Institute of Quantum Computing and Software, School of Computer Science and Engineering, Sun Yat-sen University, Guangzhou 510006, China. Email: \href{mailto:lilvzh@mail.sysu.edu.cn}{\texttt{lilvzh@mail.sysu.edu.cn}}.}
  \footnotetext[3]{\raggedright College of Computer and Data Science, Fuzhou University, Fuzhou, China. Email: \href{mailto:yezekun@fzu.edu.cn}{\texttt{yezekun@fzu.edu.cn}}.}
  \footnotetext[4]{\raggedright Quantum Science Center of Guangdong-Hong Kong-Macao Greater Bay Area, Shenzhen, China. Email: \href{mailto:zhengshenggen@quantumsc.cn}{\texttt{zhengshenggen@quantumsc.cn}}.}
  \vspace{0.35in}

  \begin{minipage}{0.88\textwidth}
  \begin{abstract}\normalsize
  A function $f:\cD\subseteq A^N\to\{0,1\}$ is \emph{permutation-invariant} if every permutation of its input positions preserves both its domain and its value. We study how this symmetry limits quantum speedups. For $1/6\leq\varepsilon<1/2$, we prove that every such function over an arbitrary finite alphabet $A$ satisfies $R_\varepsilon(f)=O(Q_\varepsilon(f)^2)$, where $R_\varepsilon$ and $Q_\varepsilon$ denote randomized and quantum query complexities with error at most $\varepsilon$. This improves the previous cubic bound (\hyperlink{cite.Chailloux2019}{ITCS 2019}).

We further establish bounds for graph and hypergraph properties, for which vertex relabeling preserves both the promise and the answer. For the same range of $\varepsilon$, we obtain $R_\varepsilon(f)=O(Q_\varepsilon(f)^4)$ for graph properties in the adjacency-matrix model and $R_\varepsilon(f)=O_k(Q_\varepsilon(f)^{2k})$ for $k$-uniform hypergraph properties with fixed $k\geq2$, where the implicit constant depends only on $k$. These improve the previous bounds $R_{1/3}(f)=O(Q_{1/3}(f)^6)$ and $R_{1/3}(f)=O_k(Q_{1/3}(f)^{3k})$, respectively (\hyperlink{cite.BenDavidEtAl2024}{SICOMP 2024}).

These bounds are tight for permutation-invariant functions and nearly tight for graph and hypergraph properties. Unstructured search gives a matching quadratic separation with $R_{1/3}(f)=\Theta\!\left(Q_{1/3}(f)^2\right)$. For every fixed $k\geq 2$, we construct explicit families of invariant promise problems on $k$-uniform hypergraphs satisfying $R_{1/3}(f)\geq Q_{1/3}(f)^{2k-o(1)}$.

  \end{abstract}
  \end{minipage}

\end{titlepage}
\renewcommand{\thefootnote}{\arabic{footnote}}
\setcounter{footnote}{0}

\begingroup
\small
\tableofcontents
\endgroup
\clearpage
\pagenumbering{arabic}
\hypersetup{pageanchor=true}

\section{Introduction}

Understanding the advantages of quantum algorithms over classical algorithms is a central question in quantum computing. Query complexity provides a setting in which these advantages can be proved without computational assumptions: an algorithm accesses its input through an oracle, and its cost is the number of queries. The deterministic query complexity $D(f)$ is the minimum number of queries needed in the worst case by a deterministic algorithm that computes $f$ on every input. The randomized query complexity $R(f)$ is the minimum number of queries needed in the worst case by a randomized algorithm that computes $f$ with error probability at most $1/3$ on every input. The quantum query complexity $Q(f)$ is defined analogously for quantum algorithms.

For total functions, quantum speedups are at most polynomial. Beals, Buhrman, Cleve, Mosca, and de Wolf proved that $D(f)$ and $Q(f)$ are polynomially related \cite{BealsEtAl2001}, and the best general bound for total Boolean functions is now $D(f)=O(Q(f)^4)$ \cite{AaronsonEtAl2021}. Since $R(f)\leq D(f)$, randomized and quantum query complexities are polynomially related as well. Thus, an exponential quantum advantage is possible only for partial functions, whose inputs are restricted by a promise. Such advantages do occur: Simon's problem gives an exponential separation \cite{Simon1997}, while the $k$-Forrelation problem yields optimal separations between randomized and quantum query complexities \cite{BansalSinha2021,SherstovEtAl2023}. In these examples, the quantum algorithms exploit specific structure imposed by the promise, such as the hidden periodicity in Simon's problem. This raises the question of how much structure is needed for large quantum speedups \cite{AaronsonAmbainis2014}.

Permutation invariance captures a notion of an unstructured problem: arbitrary permutations of the input positions preserve both the promise and the function value. Symmetric Boolean functions form an important subclass. We impose no restriction on the finite alphabet and do not require invariance under relabeling its symbols (\cref{def:permutation-invariant}).

For the study of permutation invariance, the history begins with Watrous's 2002 conjecture that randomized and quantum query complexities of are polynomially related for partial functions invariant under permutations of both the input positions and the alphabet symbols (see \cite[Conjecture~1.4]{AaronsonAmbainis2014}). Aaronson and Ambainis proved this conjecture for alphabets of size at least the input length \cite{AaronsonAmbainis2014}. Chailloux established the cubic bound over arbitrary finite alphabets, assuming invariance only under permutations of the input positions \cite{Chailloux2019}.

The study of permutation invariance also extends to graph properties. A \emph{graph property} is a Boolean function defined on a subset of  graphs on a fixed vertex set. Both its domain and its value are invariant under vertex relabeling. In the adjacency-matrix model, a query asks whether a given pair of vertices is an edge. A $k$-uniform hypergraph property is defined analogously, with each query asking whether a given set of $k$ vertices is a hyperedge. Vertex relabelings induce permutations of the edge positions, giving another natural setting for studying the effect of symmetry on quantum speedups. The case $k=2$ recovers graph properties (see \cref{sec:graph-model} for the formal definitions). Ben-David, Childs, Gily{\'e}n, Kretschmer, Podder, and Wang \cite{BenDavidEtAl2024} proved $R(f)=O(Q(f)^6)$ for graph properties and $R(f)=O_k(Q(f)^{3k})$ for $k$-uniform hypergraph properties.

We focus on the following two questions.

\begin{problem}[Permutation-invariant functions]\label{prob:strings}
What is the largest possible separation between $R(f)$ and $Q(f)$ for permutation-invariant functions?
\end{problem}

\begin{problem}[Graph and hypergraph properties]\label{prob:hypergraphs}
What is the largest possible separation between $R(f)$ and $Q(f)$ for graph and hypergraph properties?
\end{problem}

\subsection{Our Results}

Let $Q_\eps(f)$ and $R_\eps(f)$ denote the worst-case quantum and randomized query complexities of $f$ with error at most $\eps$ on every input. The notation $O_k(\cdot)$ indicates that the implicit constants depend only on $k$.

\begin{theorem}[Permutation-invariant functions over arbitrary alphabets]\label{thm:string-main}
For every permutation-invariant function $f:\cD\subseteq A^N\to\{0,1\}$ over arbitrary finite alphabet $A$ and $1/6\leq\eps<1/2$, there is a randomized algorithm that computes $f$ with error at most $\eps$ using $O(Q_\eps(f)^2)$ queries.
\end{theorem}

\begin{theorem}[Graph and hypergraph properties]\label{thm:hypergraph-main}
For every fixed integer $k\geq2$, $k$-uniform hypergraph property $f$, and $1/6\leq\eps<1/2$, there is a randomized algorithm that computes $f$ with error at most $\eps$ on every promised input using $O_k(Q_\eps(f)^{2k})$ queries.
\end{theorem}

The randomized algorithms in \cref{thm:string-main,thm:hypergraph-main} can be nonadaptive. For example, in \cref{thm:string-main}, for every $T$-query quantum algorithm, there is a randomized algorithm using $O(T^2)$ queries that achieves the same success probability as the quantum algorithm.

\begin{theorem}[Hypergraph separations]\label{thm:lower-main}
For every fixed integer $k\geq2$, there is an explicit family $\{F_N\}$ of $k$-uniform hypergraph properties on $N$ vertices such that $$ R_{1/3}(F_N)\geq N^{k-o(1)}, \qquad Q_{1/3}(F_N)\leq N^{1/2+o(1)}. $$
\end{theorem}
\cref{thm:lower-main} shows that the upper bound in \cref{thm:hypergraph-main} is nearly tight. The summary of our results is shown as Table~\ref{tab:query-bounds}.

\begin{table}[H]
\centering
\small
\renewcommand{\arraystretch}{1.2}
\setlength{\tabcolsep}{4pt}
\caption{Randomized and quantum query bounds for invariant promise problems at error $1/3$.}
\label{tab:query-bounds}
\begin{tabular}{lccc}
\toprule
Input representation & Previous upper bound & Our upper bound & Separating families\\
\midrule
Arbitrary-alphabet strings
  & \makecell[c]{$R(f)=O(Q(f)^3)$\\\cite{Chailloux2019}}
  & $R(f)=O(Q(f)^2)$
  & \makecell[c]{$R(f)=\Omega(Q(f)^2)$\\\cite{Grover1996,BennettEtAl1997}}\\
\addlinespace[6pt]
Graph adjacency matrices
  & \makecell[c]{$R(f)=O(Q(f)^6)$\\\cite{BenDavidEtAl2024}}
  & $R(f)=O(Q(f)^4)$
  & \makecell[c]{$R(f)\geq Q(f)^{4-o(1)}$\\(this work)}\\
\addlinespace[6pt]
Fixed $k$-uniform hypergraphs
  & \makecell[c]{$R(f)=O_k(Q(f)^{3k})$\\\cite{BenDavidEtAl2024}}
  & $R(f)=O_k(Q(f)^{2k})$
  & \makecell[c]{$R(f)\geq Q(f)^{2k-o(1)}$\\(this work)}\\
\bottomrule
\end{tabular}
\end{table}

\subsection{Proof Techniques}
\label{sec:proof-overview}

Our upper bounds use a small random sample to approximate a quantum computation, then represent the remaining error by another computation with the same query budget. For the lower bounds, we encode hard promise problems into hypergraphs whose structure allows quantum algorithms to recover individual input positions efficiently.

\paragraph{From small-range functions to compressed permutations.}
The earlier cubic simulation replaces a random permutation by a random function with small range \cite{Chailloux2019,Zhandry2015}. Our approach instead keeps the injectivity constraints in a coherent oracle database. It builds on the compressed-oracle method introduced by Zhandry for random functions \cite{Zhandry2019}, the compressed permutation oracle of Carolan \cite{Carolan2026}, and the improved soundness theorem of Carolan and Majenz \cite{CarolanMajenz2026}. The two additional ingredients we prove are an exact norm formula for restricting the database range to a random sample and a residual expansion that preserves the number of queries. Together, these steps give a classical simulation with the required error guarantee.

Consider a $T$-query algorithm on a uniformly relabeled string $x^\pi$, where $x^\pi_i=x_{\pi(i)}$. Each query computes $\pi(i)$, queries $x$, and uncomputes the address. The resulting circuit has $T$ queries  and $q=2T$ auxiliary permutation queries. Only the $T$ queries to the input oracle are charged. The auxiliary queries are used to analyze how much relabeling information the circuit accesses. The compressed oracle represents that information through a coherent database of partial injections $D$ from $[N]$ to $[N]$. Compression by itself does not provide a small classical sample, because the database range can contain any label in $[N]$.

\paragraph{An exact norm formula for restriction to a sample.}
We next restrict the database range to a uniformly random $s$-element set $S\subseteq[N]$. An isometric embedding places the full and sampled computations in a common space carrying a sample register. We compare compressing before this embedding with applying sampled compression after it. The operator norm of their difference bounds the state error introduced by this replacement for any unit input state.

The difference arises from extension amplitudes whose sum over all admissible labels is zero. Restricting these amplitudes to a sample can destroy this cancellation. The variance of the sampled sum determines the error.

More precisely, on a compression block based at a partial injection of size $t<s<N$, the squared operator norm of this difference is exactly $$ \frac{2(N-s)}{(s-t)(N-t-1)}. $$ Each compressed query consists of a database lookup between two compression operations. The lookup commutes with the embedding, so the two compression errors control the restriction error of the query. Combining these bounds with the compressed-permutation soundness estimate gives a total state error of $O(q/\sqrt{s})=O(K^{-1/2})$ over $q$ auxiliary permutation queries when $s=Kq^2<N$ and $K$ is a sufficiently large constant. Since $q=2T$, a sufficiently small constant error therefore requires only $O(T^2)$ sampled positions.

For a fixed sample $S$, every query in the sampled circuit can be answered from $x[S]$ and the prescribed response at $\bot$. Here $\bot\notin[N]$ is a dummy address at which the input oracle leaves the answer register unchanged. Its output expectation can therefore be evaluated by unrestricted classical postprocessing. The sampled oracle retains a coherent database of partial injections, including their injectivity constraints. It is not the evaluation oracle of a fixed small-range function.

\paragraph{A residual expansion that preserves small bias.}
To handle success probability $1/2+\beta$, we need more than a constant additive approximation. Let $h_{\cA}(x)\in[-1,1]$ be the signed output expectation of a symmetrized $T$-query algorithm. We construct a bounded postprocessing function $a_{\cA}(S,x[S])$ and another symmetrized algorithm $\cR(\cA)$ such that $$ h_{\cA}(x)=\E_S a_{\cA}(S,x[S]) +\lambda h_{\cR(\cA)}(x), \qquad \lambda=\frac18. $$ This identity represents the remaining error as a fixed multiple of another algorithm's signed output expectation. The crucial point is that the residual algorithm uses at most $T$ input queries.

To construct it, we first expand the final-state difference into a sum of terms, each containing one error operator for an auxiliary query. Each error operator is input-independent. After normalization by its norm bound, it has norm at most one and can be implemented as a component of a larger unitary circuit without postselection. We insert this operator between an ideal prefix and a sampled suffix. The ideal prefix and sampled suffix together use exactly $T$ input queries. Controlled state preparation represents each resulting cross term as a signed output expectation. Its two branches share the same $T$ input-query slots. An input-independent randomized choice among these constructions gives the residual algorithm with at most $T$ input queries. We can therefore apply the same sampling argument recursively.

Iterating the identity gives a geometric series of postprocessing functions with a residual bounded by $\lambda^L$. The sample $S$ has the same distribution in every term. By linearity of expectation, we may therefore use a single sample $S$ and evaluate all terms from $x[S]$. After a fixed normalization, choosing $L=O(\log(1/\beta))$ retains a constant fraction of the original bias without increasing the sample size. Three independent repetitions restore the original error throughout $1/6\leq\eps<1/2$. Only the postprocessing and its precision depend on the number of retained terms.

\paragraph{Sampling vertices and querying the induced hypergraph.}
We extend the sampling and residual arguments to $k$-uniform hypergraph properties to prove \cref{thm:hypergraph-main}. Each hyperedge query relabels $k$ vertex arguments using the same permutation, makes one query, and then reverses the relabeling. A $T$-query algorithm therefore uses $q=2kT$ auxiliary permutation queries. We sample $O_k(T^2)$ vertices and read the induced hypergraph, using $O_k(T^{2k})$ queries. The same single-permutation estimates and residual construction preserve the original success guarantee. Taking $k=2$ gives the quartic bound for graph properties.

\paragraph{Encoding near-maximal separations in hypergraphs.}
For the lower bound, we encode an arbitrary $m$-bit promise problem into an uncolored, connected, simple $k$-uniform hypergraph on $N=\Theta_k(m^{1/k})$ vertices. The difficulty is to locate the vertices needed to read a requested source bit after an arbitrary vertex relabeling. The general encoding in \cite[Proposition~5.1]{BenDavidEtAl2024} stores position labels in a larger alphabet. We encode these labels within the hypergraph itself and recover each requested data vertex coherently using $\widetilde O_k(\sqrt N)$ queries.

The construction has two vertex classes: data vertices with distinct binary addresses and a logarithmic number of address vertices. The source bits are stored in hyperedges among a fixed fraction of the data vertices. The address vertices have higher degrees, and a rigid tree determines their order. Hyperedges containing one address vertex record the OR of the corresponding address bits of their data vertices.

Each hyperedge query to the canonically labeled encoding can be answered using at most one source query, preserving randomized lower bounds. For the quantum reduction, degree estimates and amplitude amplification \cite{Grover1996,BrassardEtAl2002} identify the address vertices. For each address coordinate, a query returning zero identifies data vertices whose corresponding bits are all zero. Using these vertices to fill the remaining positions of a hyperedge query then reads that bit of any data vertex exactly. Amplitude amplification can therefore locate each requested data vertex coherently.

We apply this encoding to the $\ell$-Forrelation problems of Bansal and Sinha~\cite{BansalSinha2021}, choosing the order $\ell$ to increase slowly with the input size. These source problems require $m^{1-o(1)}$ randomized queries but only polylogarithmically many quantum queries. The encoding transfers this separation to hypergraph properties and yields $$ R_{1/3}(F_N)\geq N^{k-o(1)}, \qquad Q_{1/3}(F_N)\leq N^{1/2+o(1)}. $$ This proves \cref{thm:lower-main} using connected simple $k$-uniform hypergraphs without vertex colors or an additional address oracle.

\paragraph{AI assistance.}
We used OpenAI ChatGPT (Codex) to assist with the development and review of the sampling argument, residual expansion, hypergraph simulation, and hypergraph encoding in \cref{sec:string-sampling,sec:small-bias,sec:hypergraph-simulation,sec:lower-bound}. Further details are given in the AI Methodology statement at the end of the paper.

\subsection{Organization}

\Cref{sec:preliminaries} introduces the query models and notations. \Cref{sec:string-sampling} develops a quadratic simulation for strings at constant error. \Cref{sec:small-bias} preserves the original success guarantee even when the success bias is arbitrarily small. \Cref{sec:hypergraph-simulation} extends the simulator to graph and hypergraph properties and proves \cref{thm:hypergraph-main}. \Cref{sec:lower-bound} establishes the hypergraph encoding and uses it to prove \cref{thm:lower-main}.

\section{Preliminaries}\label{sec:preliminaries}

We first specify the oracle and error conventions, then define the symmetries used in the simulation and separation results.

\subsection{Query Model}

Let $[N]=\{1,\ldots,N\}$ and $f:\cD\subseteq A^N\to\{0,1\}$ be a function on strings over a finite alphabet $A$. Its domain $\cD$ is the \emph{promise}. The function is \emph{total} when $\cD=A^N$. Unless stated otherwise, the functions considered in this paper may be total or partial. We use \emph{Boolean function} only when the input alphabet is $\{0,1\}$. Fix a known injective encoding $\mathrm{enc}_A:A\to\{0,1\}^b$. The input oracle acts as $$ O_x\ket{i,z}=\ket{i,z\oplus\mathrm{enc}_A(x_i)}, \qquad i\in[N]. $$ We identify alphabet symbols with their encodings when no confusion arises. Each application of $O_x$ counts as one query, regardless of the encoding length $b$. The proofs also use auxiliary permutation queries to implement random relabelings. These operations are independent of the unknown input and do not contribute to its query complexity. Unless explicitly called an auxiliary permutation query, a query means a call to the input oracle. In the classical model, this is a request for one input entry. Algorithms may use arbitrary finite workspaces and arbitrary input-independent computation. For $0\leq\eps<1/2$, the \emph{quantum query complexity} $Q_\eps(f)$ is the minimum number of queries required by a quantum algorithm to compute $f$ with error at most $\eps$ on every input in its domain. The \emph{randomized query complexity} $R_\eps(f)$ is defined analogously using randomized algorithms. In both definitions, the number of queries is measured in the worst case. We write $Q(f)=Q_{1/3}(f)$ and $R(f)=R_{1/3}(f)$. A randomized algorithm is \emph{nonadaptive} if, for each choice of its internal randomness, all query addresses are fixed before any input values are observed.

Any algorithm making at most $T$ queries can be modified to make exactly $T$ queries without changing its output distribution. We also allow a dummy address $\bot$ at which the oracle acts as the identity.

We use $\norm v$ for the Euclidean norm and $\norm B=\sup_{\norm v=1}\norm{Bv}$ for the operator norm. An operator $B$ is a \emph{contraction} if $\norm B\leq1$ and an \emph{isometry} if $B^\dagger B=I$.

To express success probabilities in a form suited to the simulation, we use signed outputs. For an algorithm $\cA$ with output $B_x\in\{0,1\}$, let $Y_x=2B_x-1\in\{-1,+1\}$ be its signed output. After purifying randomness and intermediate measurements, there are a final pure state $\ket{\psi_x}$ and an input-independent Hermitian contraction $M$ such that $$ h_{\cA}(x):=\E[Y_x] =\bra{\psi_x}M\ket{\psi_x}, \qquad M=M^\dagger,\quad\norm{M}\leq1. $$ The two-outcome measurement with effects $(I+M)/2$ and $(I-M)/2$ produces a signed output with expectation $h_{\cA}(x)$. Let $F(x)=2f(x)-1$. Then \begin{equation}\label{eq:signed-success} \Pr[\cA(x)=f(x)] =\frac{1+F(x)h_{\cA}(x)}{2}. \end{equation} An algorithm has worst-case success bias at least $\beta>0$ if it succeeds with probability at least $1/2+\beta$ on every promised input. By \eqref{eq:signed-success}, this condition is equivalent to $$ F(x)h_{\cA}(x)\geq2\beta \qquad\text{for all }x\in\cD. $$

\subsection{Permutation Invariance and Symmetrization}

Permutation invariance allows us to average an algorithm over input relabelings without reducing its worst-case success guarantee. We formalize this symmetry and introduce the classical sampling model used to approximate symmetrized output expectations.

\begin{definition}[Permutation-invariant functions]\label{def:permutation-invariant}
Let $S_N$ denote the symmetric group on $[N]$, let $A$ be a finite alphabet, and let $f:\cD\subseteq A^N\to\{0,1\}$ be a function. For $\pi\in S_N$, write $x^\pi_i=x_{\pi(i)}$. The function $f$ is \emph{permutation-invariant} if, for every $x\in\cD$ and $\pi\in S_N$, $$ x^\pi\in\cD \qquad\text{and}\qquad f(x^\pi)=f(x). $$
\end{definition}

For any algorithm $\cA$, define its symmetrized signed expectation by $$ \bar h_{\cA}(x)=\E_{\pi\in S_N}h_{\cA}(x^\pi). $$ If $\cA$ has worst-case success bias at least $\beta$ on an invariant promise, then $F(x)\bar h_{\cA}(x)\geq2\beta$ on every promised input. We may purify the uniform choice of $\pi$ into an external permutation register, which the original circuit accesses only through the compiled relabeling queries.

We will approximate these symmetrized expectations using classical simulators that read a random sample. Such a simulator chooses a uniform $s$-element subset $S\subseteq[N]$ before observing any input values and queries the labeled restriction $x[S]=(x_i)_{i\in S}$. It then evaluates a known postprocessing function $a(S,x[S])\in[-1,1]$, with no further queries. Outputting $+1$ with probability $(1+a(S,x[S]))/2$ and $-1$ otherwise gives signed expectation $$ \E_{S:|S|=s}a(S,x[S]). $$ The simulator is nonadaptive and uses $s$ queries, regardless of the computation required to evaluate the postprocessing function.

\subsection{Graph and Hypergraph Properties}\label{sec:graph-model}

For graph and hypergraph properties, vertex relabelings induce permutations of the queried edge positions. For an integer $k\geq2$ and a finite set $V$, let $\binom Vk$ denote its set of $k$-element subsets. A simple $k$-uniform hypergraph on $V$ is specified by a function $H:\binom Vk\to\{0,1\}$, with $H(e)=1$ exactly when $e$ is a hyperedge. The case $k=2$ is a simple undirected graph. For $V=[N]$ and $\pi\in S_N$, define $\pi(e)=\{\pi(v):v\in e\}$ and $H^\pi(e)=H(\pi(e))$.

\begin{definition}[Graph and hypergraph properties]\label{def:hypergraph-property}
A \emph{$k$-uniform hypergraph property} is a function $$ f:\cD\subseteq\{0,1\}^{\binom{[N]}k}\to\{0,1\} $$ such that, for every $H\in\cD$ and $\pi\in S_N$, $$ H^\pi\in\cD \qquad\text{and}\qquad f(H^\pi)=f(H). $$ A \emph{graph property} is the case $k=2$. The property is total when $\cD=\{0,1\}^{\binom{[N]}k}$.
\end{definition}

Unless stated otherwise, graph and hypergraph properties may have promises. For oracle access, extend $H$ to ordered vertex tuples by $$ H(v_1,\ldots,v_k)= \begin{cases} H(\{v_1,\ldots,v_k\}), & \text{if the vertices are distinct},\\ 0, & \text{otherwise}. \end{cases} $$ The query oracle XORs this value into an answer bit. For $k=2$, this is the adjacency-matrix model, with a symmetric matrix and zero diagonal. All hypergraphs considered here are uncolored: the input specifies only hyperedge membership, with no vertex colors. The incidence graph has one node for each vertex and each hyperedge, and joins a vertex to the hyperedges containing it. A hypergraph is connected if its incidence graph is connected.

\section{Quadratic Simulator for Permutation-Invariant Functions}\label{sec:string-sampling}

Fix an algorithm $\cA$ using at most $T\geq1$ queries. Using the convention from \cref{sec:preliminaries}, we arrange for exactly $T$ queries without changing its output distribution and let $q=2T$. Our goal is to construct a uniform sample $S$ of size $O(T^2)$ and a function $a_{\cA}(S,x[S])\in[-1,1]$ such that $$ \left| \bar h_{\cA}(x)-\E_S a_{\cA}(S,x[S]) \right| \leq\frac18 $$ for every input $x$. The function uses no input values outside the sample. This gives the constant-error quadratic simulation in \cref{cor:constant-error-simulation}.

After compiling the relabeling, the proof has four steps. We apply the compressed-permutation estimate (\cref{thm:CM}), prove an exact norm formula for restriction to a sample (\cref{lem:exact-norm}), combine the two approximations (\cref{lem:combined-approximation}), and show that the sampled circuit uses only the sampled input (\cref{lem:local-approximation}). All $T$ queries remain in the comparison. Only the $q$ auxiliary permutation queries are replaced.

Table~\ref{tab:relabeling-representations} summarizes the three representations of the relabeling information used below.

\begin{table}[H]
\centering
\small
\setlength{\tabcolsep}{6pt}
\caption{Representations of the relabeling information.}
\label{tab:relabeling-representations}
\begin{tabular}{lcc}
\toprule
Representation & Space & Map into this representation\\
\midrule
Complete permutation $\pi$
  & $\cP_N$ & ---\\[4pt]
Partial injection $D$
  & $\cL_N$ & $V_N:\cP_N\to\cL_N$\\[4pt]
\makecell[l]{Sample $S$ and database $D$\\with $\im D\subseteq S$}
  & $\displaystyle\bigoplus_{|S|=s}\ket S\otimes\cL_S$
  & $J_s$ from $\cL_{N,\leq s}$\\
\bottomrule
\end{tabular}
\end{table}

On the corresponding low-degree permutation subspaces, $W=J_sV_N$ combines the two maps.

\subsection{Compiled Relabeling Queries}
\label{sec:compiled-relabeling}

To query $x^\pi_i$, the circuit first computes $\pi(i)$ into an auxiliary address register initialized to $\ket\bot$. It then applies $O_x$ once using this address register and uncomputes $\pi(i)$ to restore the register to $\ket\bot$. Both the computation and its reversal use the oracle for $\pi$ without querying $\pi^{-1}$.

The circuit retains this register as part of its workspace and uses a new register initialized to $\ket\bot$ for the next query. The resulting circuit implements the original algorithm using $T$ queries to $O_x$ and $q=2T$ auxiliary permutation queries.

On the register spanned by $\{\bot\}\cup[N]$, let $S_y$ swap $\ket\bot$ and $\ket y$. The purified auxiliary permutation query at address $i\in[N]$ is $$ P_i=\sum_{y\in[N]}S_y\otimes \sum_{\pi:\pi(i)=y}\ket\pi\!\bra\pi. $$ For a query controlled on the dummy input address, set $P_\bot=I$ and take its compressed and sampled counterparts to be the identity. Let $\cP_N=\operatorname{span}\{\ket\pi:\pi\in S_N\}$ be the Hilbert space of a register storing the permutation. This register is initialized to $\ket{\mathrm{unif}_N}=N!^{-1/2}\sum_{\pi\in S_N}\ket\pi$. Tracing out this register recovers the average over a uniformly random permutation.

\subsection{Compressed Permutations}

The compiled circuit stores a complete random permutation. We compare its auxiliary queries with a compressed oracle whose basis states are partial injections.

A database is a partial injection $D$, viewed as an injective map $D:\dom(D)\to[N]$ with $\dom(D)\subseteq[N]$. We write $|D|=|\dom D|=|\im D|$ for its number of records. Let $\cL_N$ be the Hilbert space with these databases as an orthonormal basis. Let $\cL_{N,\leq t}$ and $\cL_{N,t}$ denote the spans of database basis states of size at most $t$ and exactly $t$, respectively. Fix an address $i$. Every database $H$ with $i\notin\dom H$ defines the block consisting of $\ket H$ and all $\ket{H\cup\{i\mapsto y\}}$ with $y\notin\im H$. Define $$ \ket{u^N_{i,H}}= \frac1{\sqrt{N-|H|}} \sum_{y\notin\im H}\ket{H\cup\{i\mapsto y\}}. $$ The unitary $C_i^N$ swaps $\ket H$ and $\ket{u^N_{i,H}}$, fixing the orthogonal complement of their span in each block. If $Q_{i,y}$ projects onto databases with $D(i)=y$ and $Q_{i,\bot}$ onto those without $i$, define $$ L_i^N=I\otimes Q_{i,\bot} +\sum_{y\in[N]}S_y\otimes Q_{i,y}, \qquad \cO_i^N=C_i^NL_i^NC_i^N. $$

For a partial injection $D$, its completion state is the uniform superposition of all permutations extending $D$. Let $\cH_{\leq t}$ be the span of completion states with $|D|\leq t$, and let $\Pi_{\leq t}$ be the corresponding orthogonal projector. With $\cH_{\leq-1}=\{0\}$, define $$ \cH_t=\cH_{\leq t}\cap(\cH_{\leq t-1})^\perp. $$ One purified permutation query raises this degree by at most one. The following theorem provides an isometry from permutation states to database states and bounds the error in replacing one purified query by a compressed query in the swap-oracle model defined above.

\begin{theorem}[Carolan--Majenz, Theorem 4.2, Eq.~(12) \cite{CarolanMajenz2026}]\label{thm:CM}
For $N\geq3$ there is an input-independent isometry $V_N:\cP_N\to\cL_N$ such that $V_N\cH_t\subseteq\cL_{N,t}$ and $V_N\ket{\mathrm{unif}_N}=\ket\varnothing$. If $3t+2\leq N$, then \begin{equation}\label{eq:CM} \norm{(\cO_i^NV_N-V_NP_i)\Pi_{\leq t}} \leq\frac{60}{\sqrt{N-t}}+\frac{t+1}{N-2t-1}. \end{equation}
\end{theorem}

\subsection{Restricting the Database Range}

A small database may still contain labels anywhere in $[N]$. To obtain a simulator that reads only a sample, we restrict the database range and bound the change in the compressed queries.

Choose $q+1<s<N$. For each $s$-set $S\subseteq[N]$, let $\cL_S$ be spanned by databases with image contained in $S$. Define $C_i^S$, $L_i^S$, and $\cO_i^S$ as above, restricting extension labels to $S$. When $|H|=s$ and $i\notin\dom H$, the block has no extension. Define $C_i^S$ to be the identity on this one-dimensional block. Thus these operators are defined and unitary on all of $\cL_S$.

Define an isometry only on databases of size at most $s$: \begin{equation}\label{eq:sample-isometry} J_s:\cL_{N,\leq s}\longrightarrow \bigoplus_{S:|S|=s}\ket S\otimes\cL_S, \qquad J_s\ket D= \binom{N-|D|}{s-|D|}^{-1/2} \sum_{\substack{|S|=s\\\im D\subseteq S}}\ket S\ket D. \end{equation} Different databases have orthogonal images, and each image has norm one. We use hats to denote direct sums over all subsets $S\subseteq[N]$ of size $s$. For example, $$ \widehat C_i=\bigoplus_{|S|=s}C_i^S. $$ The operator $L_i^N$ leaves the database register unchanged. Consequently, $\widehat L_iJ_s=J_sL_i^N$ on the domain of $J_s$. The remaining error comes from the two possible orders of compression and restriction to a sample. The next lemma computes its norm exactly on each compression block.

\begin{lemma}[Exact norm of the restriction error]\label{lem:exact-norm}
On the block based at $H$ with $i\notin\dom H$ and $t=|H|<s<N$, \begin{equation}\label{eq:exact-norm} \norm{\widehat C_iJ_s-J_sC_i^N}^2 =\frac{2(N-s)}{(s-t)(N-t-1)}. \end{equation}
\end{lemma}

\begin{proof}
We decompose the compression block into the span of its two swapped states and the orthogonal complement. The two orders of compression and restriction agree on the first subspace; on the second, we compute their difference by summing over sample blocks.

Let $a=N-t$, $b=s-t$, $d_0=\binom ab$, and $d_1=\binom{a-1}{b-1}$. Identify $[N]\setminus\im H$ with $[a]$, and write $\ket0=\ket H$ and $\ket y=\ket{H\cup\{i\mapsto y\}}$ for $y\in[a]$. Set $\ket{u_a}=a^{-1/2}\sum_y\ket y$. A sample extending $\im H$ corresponds to a $b$-set $K\subseteq[a]$. Write $\ket{u_K}=b^{-1/2}\sum_{y\in K}\ket y$. Since $ad_1=bd_0$, the components in sample block $K$ satisfy $$ (J_s\ket0)_K=\frac{\ket0}{\sqrt{d_0}}, \qquad (J_s\ket{u_a})_K=\frac{\ket{u_K}}{\sqrt{d_0}}. $$ The full and restricted compressions swap the corresponding two vectors. Thus the restriction $E_H$ of $\widehat C_iJ_s-J_sC_i^N$ to this block vanishes on $\operatorname{span}\{\ket0,\ket{u_a}\}$.

It remains to consider the orthogonal complement. For $v=\sum_yv_y\ket y$ with $\sum_yv_y=0$, the full compression fixes $v$, whereas the restricted compression swaps its component along $\ket{u_K}$ with $\ket0$. Consequently, $$ (E_Hv)_K=\frac{\sum_{y\in K}v_y}{\sqrt{bd_1}}(\ket0-\ket{u_K}). $$ Each label belongs to $d_1$ of the $b$-sets $K$, and each pair of distinct labels belongs to $d_1(b-1)/(a-1)$ of them. Summing over the orthogonal sample blocks and using $\sum_yv_y=0$ gives \begin{align*} \norm{E_Hv}^2 &=\frac2b\left(\sum_y|v_y|^2+ \frac{b-1}{a-1}\sum_{y\ne z}v_y\overline{v_z}\right)\\ &=\frac{2(a-b)}{b(a-1)}\norm v^2. \end{align*} Every vector in the block is the orthogonal sum of a vector in $\operatorname{span}\{\ket0,\ket{u_a}\}$ and such a zero-sum vector $v$. Since $a>b\geq1$, nonzero zero-sum vectors exist, so $$ \norm{E_H}^2=\frac{2(a-b)}{b(a-1)} =\frac{2(N-s)}{(s-t)(N-t-1)}. $$
\end{proof}

For a fixed address $i$, distinct compression blocks remain orthogonal under $J_s$ and are preserved by compression. Every database of size at most $t$ lies in a block whose base has size at most $t$. Thus \cref{lem:exact-norm}, with $P_{\leq t}^{\cL}$ denoting the projector onto $\cL_{N,\leq t}$, gives \begin{equation}\label{eq:compression-diff} \norm{(\widehat C_iJ_s-J_sC_i^N)P_{\leq t}^{\cL}} \leq \sqrt{\frac{2}{s-t}}, \qquad t\leq s-1. \end{equation} The same bound holds for coherently controlled addresses, since the operator is block diagonal in the address register.

Let $E_i=\widehat C_iJ_s-J_sC_i^N$ wherever these compositions are defined. Using $\widehat L_iJ_s=J_sL_i^N$, we obtain $$ \widehat{\cO}_iJ_s-J_s\cO_i^N =\widehat C_i\widehat L_iE_i+E_iL_i^NC_i^N. $$ The operator $C_i^N$ increases database size by at most one, and $L_i^N$ leaves it unchanged. For $t\leq s-2$, all compositions are therefore defined on $\cL_{N,\leq t}$. Unitarity and \eqref{eq:compression-diff} at sizes $t$ and $t+1$ give \begin{equation}\label{eq:sample-query-error} \begin{aligned} \norm{(\widehat{\cO}_iJ_s-J_s\cO_i^N)P_{\leq t}^{\cL}} &\leq \sqrt{\frac{2}{s-t}}+\sqrt{\frac{2}{s-t-1}}\\ &\leq 2\sqrt{\frac{2}{s-t-1}},\qquad t\leq s-2. \end{aligned} \end{equation}

\subsection{Approximating Permutation Queries}

We now combine the compressed-permutation approximation with the estimate for restriction to a sample to bound the error in replacing one ideal permutation query by a sampled query.

\begin{lemma}[Combined query approximation]
\label{lem:combined-approximation}
Let $q\geq1$, $q+1<s<N$, and $3q+2\leq N$. The map $W=J_sV_N$ is an isometry on $\cH_{\leq q+1}$. For every $0\leq t\leq q$, \begin{equation}\label{eq:eta} \norm{(\widehat{\cO}_iW-WP_i)\Pi_{\leq t}} \leq \eta_{N,s,q} := \frac{60}{\sqrt{N-q}} +\frac{q+1}{N-2q-1} +2\sqrt{\frac{2}{s-q-1}}. \end{equation}
\end{lemma}

\begin{proof}
We compare ideal and sampled queries through the full compressed oracle, applying the two preceding estimates in turn.

By \cref{thm:CM}, $V_N\cH_{\leq q+1}\subseteq\cL_{N,\leq q+1}$. Since $q+1<s$, the map $W=J_sV_N$ is well defined and isometric on $\cH_{\leq q+1}$. For $0\leq t\leq q$, insert $J_s\cO_i^NV_N$ to obtain \begin{align*} (\widehat{\cO}_iW-WP_i)\Pi_{\leq t} ={}&(\widehat{\cO}_iJ_s-J_s\cO_i^N)V_N\Pi_{\leq t}\\ &+J_s(\cO_i^NV_N-V_NP_i)\Pi_{\leq t}. \end{align*} The first term is bounded by \eqref{eq:sample-query-error}. In the second term, $\cO_i^N=C_i^NL_i^NC_i^N$ increases database size by at most two, while $P_i$ increases degree by at most one. Hence the argument to $J_s$ is supported on databases of size at most $t+2\leq q+2\leq s$, where $J_s$ preserves norms. Together with \eqref{eq:CM}, this gives $$ \norm{(\widehat{\cO}_iW-WP_i)\Pi_{\leq t}} \leq 2\sqrt{\frac{2}{s-t-1}}+ \frac{60}{\sqrt{N-t}}+\frac{t+1}{N-2t-1} \leq\eta_{N,s,q}. $$ For coherently controlled addresses, the operator is block diagonal in the address register, so the same estimate holds.

Extend $W$ by mapping $\cH_{\leq q+1}^{\perp}$ isometrically into an additional orthogonal target space, and extend each sampled query by the identity on that space. This preserves \eqref{eq:eta}: for $t\leq q$, both the input in $\cH_{\leq t}$ and its image under $P_i$ in $\cH_{\leq t+1}$ use only the original action of $W$. The extension acts only on external memory and is input-independent.
\end{proof}

After adding suitable ancillary registers, the extended isometry has an input-independent unitary implementation on external memory. Since $W$ acts only on external memory, it commutes with every operation of the original circuit, including the input oracle.

The scale of \eqref{eq:eta} explains the quadratic sample size. For $q\geq1$, $N\geq4q+2$, and $s\geq2(q+1)$, the explicit bound gives $$ \eta_{N,s,q} \leq\frac{60\sqrt2}{\sqrt N}+\frac{2(q+1)}{N}+\frac4{\sqrt s} =O\!\left(N^{-1/2}+q/N+s^{-1/2}\right). $$ If also $q^2\leq s<N$, the accumulated state error over $q$ auxiliary queries is $O(q\eta_{N,s,q})=O(q/\sqrt s)$. Thus a sample of size $s=Kq^2$ makes this error a sufficiently small constant when $K$ is sufficiently large.

We fix $K=10^8$ to meet the error bound required below. If $N\leq Kq^2$, query the entire string. Otherwise, set $s=Kq^2$. These parameters satisfy the conditions above and the hypotheses of \cref{lem:combined-approximation}. The displayed estimate then gives $$ 2q\eta_{N,s,q} \leq\frac{120\sqrt2+8}{\sqrt K}+\frac8K <\frac18. $$ Hence \begin{equation}\label{eq:lambda-budget} 2q\eta_{N,s,q}\leq\lambda:=\frac18. \end{equation}

\subsection{Nonadaptive Simulation at Constant Error}

We first show that the sampled circuit depends only on the sampled input and approximates the signed output expectation of the ideal circuit.

\begin{lemma}[Approximation from the sampled input]\label{lem:local-approximation}
Let $\cA$ be a $T$-query algorithm on strings in $A^N$, and let $q=2T\geq1$. Assume that $$ q+1<s<N, \qquad 3q+2\leq N. $$ There is a function $a_{\cA}(S,x[S])\in[-1,1]$ such that \begin{equation}\label{eq:local-expectation} \ell_{\cA}(x) = \E_{S:|S|=s}a_{\cA}(S,x[S]) \end{equation} satisfies $$ \abs{\bar h_{\cA}(x)-\ell_{\cA}(x)} \leq 2q\eta_{N,s,q} $$ for every input $x$.
\end{lemma}

\begin{proof}
We construct the sampled circuit and show that its restriction to each sample block depends only on $x[S]$. We then compare its final state with that of the ideal circuit using the one-query estimate.

Replace every auxiliary permutation query $P_i$ in the compiled circuit by $\widehat{\cO}_i$. Since $V_N\ket{\mathrm{unif}_N}=\ket\varnothing$, the external registers start in $$ W\ket{\mathrm{unif}_N}=\binom Ns^{-1/2} \sum_{\substack{S\subseteq[N]\\|S|=s}}\ket{S,\varnothing}. $$ Thus every sample initially has the same weight and an empty database. The sampled queries preserve the sample label because they are direct sums over $S$.

Fix a sample $S$. In this block, each auxiliary query is $\cO_i^S=C_i^SL_i^SC_i^S$. The operators $C_i^S$ act only on the database, while $L_i^S$ applies $S_y$ to the auxiliary answer register if the database records $i\mapsto y$, and otherwise acts as the identity. Every recorded value belongs to $S$. An auxiliary answer register initialized to $\ket\bot$ therefore remains supported on $$ \operatorname{span}\bigl(\{\ket\bot\}\cup\{\ket y:y\in S\}\bigr), $$ even when entangled with other registers. When this register serves as the query address, the required actions are $$ \ket y\ket z\longmapsto\ket y\ket{z\oplus x_y}\quad(y\in S), \qquad \ket\bot\ket z\longmapsto\ket\bot\ket z. $$ They are determined by $x[S]$ and the prescribed zero response at $\bot$. The second auxiliary query need not return the address register to $\ket\bot$. We retain it as workspace and use a fresh register initialized to $\ket\bot$ for the next query. The support argument thus applies at every query, so the circuit in this block depends only on $S$, $x[S]$, and the description of $\cA$.

Let $a_{\cA}(S,x[S])=\E[2B_S-1]\in[-1,1]$, where $B_S$ is the output of this circuit. All operations preserve the sample label and the output measurement acts only on accessible registers. The overall signed expectation is therefore the uniform average in \eqref{eq:local-expectation}.

It remains to bound the approximation error. Before the $j$th auxiliary query, the ideal state has external degree at most $j-1$. Operations between auxiliary queries preserve this degree and commute with $W$. The sampled suffixes are unitary. A telescoping expansion with ideal prefixes and sampled suffixes therefore bounds each of the $q$ query replacements by \eqref{eq:eta}. Since the initial sampled state is exactly the image under $W$ of the ideal state, the final states satisfy $$ \norm{u_x-v_x}\leq q\eta_{N,s,q}, $$ where $u_x$ is the ideal final state mapped by $W$ and $v_x$ is the sampled final state. The signed output observable $M$, tensored with the identity on external memory, has $\norm M\leq1$. As $W$ acts only on that memory, Cauchy--Schwarz gives \begin{align*} \abs{\bar h_{\cA}(x)-\ell_{\cA}(x)} &=\abs{\bra{u_x}M\ket{u_x}-\bra{v_x}M\ket{v_x}}\\ &\leq2\norm{u_x-v_x}\leq2q\eta_{N,s,q}. \end{align*}
\end{proof}

For $s=Kq^2<N$, \eqref{eq:lambda-budget} gives \begin{equation}\label{eq:constant-additive} \abs{\bar h_{\cA}(x)-\ell_{\cA}(x)} \leq\frac18. \end{equation}

We now apply \cref{lem:local-approximation} to a bounded-error algorithm for a permutation-invariant promise problem.

\begin{corollary}[Constant-error simulation]
\label{cor:constant-error-simulation}
Every permutation-invariant function $f$ over a finite alphabet satisfies $$ R_{1/3}(f)=O(Q_{1/3}(f)^2). $$ The randomized algorithm is nonadaptive.
\end{corollary}

\begin{proof}
Let $T=Q_{1/3}(f)$ and $q=2T$. If $T=0$, then $f$ is constant on its promise and requires no randomized queries. If $T\geq1$ and $N\leq Kq^2$, query the full input.

Otherwise, \cref{lem:local-approximation} with $s=Kq^2$ gives a sampled experiment with success probability at least $1/2+5/48$. Rounding each conditional output probability to a dyadic grid within $5/96$ gives a finite classical decision rule with success probability at least $1/2+5/96$. A constant number of independent repetitions and majority vote reduces the error to $1/3$, using $O(T^2)$ queries whose locations can all be selected in advance.
\end{proof}

\section{A Residual Expansion for Small Bias}\label{sec:small-bias}

The constant additive approximation from \eqref{eq:constant-additive} does not by itself preserve an arbitrarily small positive success bias. If the original signed expectation is only $2\beta$ in the correct direction, an additive error of $1/8$ may eliminate that guarantee.

The operator estimate \eqref{eq:eta} provides more structure than an additive bound on output expectations. The error operator for each auxiliary query is independent of the input. After division by its norm bound, it has norm at most one and can be realized as a component of a unitary circuit. Inserting this component between an ideal prefix and a sampled suffix uses at most the original $T$ queries. The controlled preparation below converts the resulting cross terms into signed output expectations without increasing this bound.

We thereby represent the difference between the ideal and sampled expectations exactly as a constant multiple of another algorithm's signed expectation. The same construction can be applied to that residual algorithm. Iteration reduces the remaining term geometrically while keeping the sample size fixed.

Fix an integer $T\geq1$ and let $$ q=2T,\qquad K=10^8,\qquad \lambda=\frac18. $$ Set $s=N$ if $N\leq\lceil Kq^2\rceil$, and set $s=\lceil Kq^2\rceil$ otherwise. Throughout the recursion, $T$ is a fixed upper bound on the number of queries. Even if a residual algorithm uses fewer queries, we continue to use this same value of $T$ when choosing $s$.

\subsection{Controlled State Preparation}

The following lemma allows us to represent an overlap between two prepared states without adding their query counts.

\begin{lemma}[Controlled state preparation]
\label{lem:aligned-preparation}
Suppose two circuits each use at most $T$ applications of the same XOR oracle $O_x$. After adding initialized registers and identifying their output spaces, let their normalized output states be $\ket{\widetilde a_x}$ and $\ket{\widetilde b_x}$. There is a circuit using at most $T$ queries that prepares $$ \frac{\ket0\ket{\widetilde a_x} + \ket1\ket{\widetilde b_x}}{\sqrt2}. $$

Let $P$ be an input-independent orthogonal projector onto a common output subspace, and let $$ \ket{a_x}=P\ket{\widetilde a_x}, \qquad \ket{b_x}=P\ket{\widetilde b_x}. $$ For every input-independent Hermitian contraction $M$ on this subspace, the quantity $$ \operatorname{Re}\bra{a_x}M\ket{b_x} $$ is the signed output expectation of an algorithm using at most $T$ queries.
\end{lemma}

\begin{proof}
We prepare the two branches using the same $T$ query positions and then recover the desired real part as a signed measurement expectation. Purify all randomness and intermediate measurements. Using the convention from \cref{sec:preliminaries}, pad both circuits with queries that leave the state unchanged so that each circuit makes exactly $T$ queries. Number these queries from $1$ to $T$.

Prepare a control qubit in $(\ket0+\ket1)/\sqrt2$. The control qubit selects which circuit's input-independent operations to apply. At the $j$th query, use controlled swaps to move the selected circuit's address and answer registers into the registers on which $O_x$ acts. Apply $O_x$ once and reverse the swaps.

For each padding query, use any valid address and an auxiliary answer register in the state $$ \ket{+_b}=2^{-b/2}\sum_{z\in\{0,1\}^b}\ket z. $$ This state is unchanged by the map $\ket z\mapsto\ket{z\oplus a}$ for every $a\in\{0,1\}^b$. Thus one ordinary query to $O_x$ implements the $j$th query in both branches. The entire controlled preparation uses $T$ queries. The same procedure implements a query at $\bot$ by routing that branch to a fixed valid address and the auxiliary answer state, then undoing the swaps.

The residual construction will also use states obtained from an input-independent contraction $B$. Embed its input and output spaces into a common finite space and extend $B$ by zero on the additional directions. The standard unitary dilation \cite[Theorem~2]{Robinson2018} gives an input-independent unitary $U_B$ with the action $$ \ket0\ket\psi \longmapsto \ket0 B\ket\psi + \ket1\sqrt{I-B^\dagger B}\ket\psi. $$ The component with flag zero is exactly $B\ket\psi$ without normalization. Retain the flag without measurement and leave it unchanged in subsequent operations, so no postselection is used. Implementing $U_B$ requires no queries.

Add ancillary registers as needed so that the desired components lie in a common output subspace, with $P$ also selecting the required flag values. Extend $M$ by zero on the orthogonal complement to obtain $\widetilde M$. Let $\ket{\Omega_x}$ be the controlled state prepared above, and let $X_c=\ket0\!\bra1+\ket1\!\bra0$ act on the control qubit. Then $$ \bra{\Omega_x}(X_c\otimes\widetilde M)\ket{\Omega_x} =\operatorname{Re}\bra{a_x}M\ket{b_x}. $$ Since $X_c\otimes\widetilde M$ is a Hermitian contraction, the measurement described in \cref{sec:preliminaries} realizes this signed expectation.
\end{proof}

\subsection{The Exact Residual Identity}

We now apply controlled state preparation to the difference between the output expectations of the ideal and sampled circuits. The next lemma expresses this difference as a fixed multiple of another algorithm's signed output expectation while preserving the original query bound.

\begin{lemma}[Residual identity with the same query bound]
\label{lem:residual}
Fix an integer $T\geq1$ and the sample size $s$ defined above. For every string algorithm $\cA$ using at most $T$ queries, there are a function $a_{\cA}(S,x[S])\in[-1,1]$ and a symmetrized algorithm $\cR(\cA)$ using at most $T$ queries such that \begin{equation}\label{eq:residual} \bar h_{\cA}(x) = \E_S a_{\cA}(S,x[S]) + \lambda h_{\cR(\cA)}(x), \qquad \lambda=\frac18, \end{equation} for every $x\in A^N$. The sample $S$ is uniform among the $s$-element subsets of $[N]$.
\end{lemma}

\begin{proof}
We first expand the state difference into local query errors, then realize the resulting cross terms within the original query bound. Symmetrizing this representation gives the claimed residual identity. We handle the full-input case before carrying out these three steps.

If $s=N$, take $a_{\cA}([N],x)=\bar h_{\cA}(x)$ and let $\cR(\cA)$ output a fair sign.

Assume $s<N$ and let $$ \eta=\eta_{N,s,q}. $$ By the choice of $s$ and \eqref{eq:lambda-budget}, $$ 2q\eta\leq\lambda. $$ This inequality will let us combine the normalized local errors into a single residual algorithm.

\paragraph{Step 1: a telescoping expansion of the state difference.}
We use the ideal and sampled circuits from \cref{sec:string-sampling}. The query sequence has been extended to exactly $T$ queries using the convention from \cref{sec:preliminaries}, so the compiled circuit has exactly $q=2T$ auxiliary permutation queries. Let $\ket{u_x}$ be the ideal final state mapped into the common space by $W$, and let $\ket{v_x}$ be the sampled final state. Both states are normalized.

Number the auxiliary permutation queries from $1$ to $q$. Let $P_j$ and $\widehat P_j$ denote the ideal and sampled operators at the $j$th such query, including coherent control of the address. Identities on unaffected registers are implicit below.

Define \begin{equation}\label{eq:Kj} K_j = \frac{ (WP_j-\widehat P_jW)\Pi_{\leq j-1} }{\eta}. \end{equation} The operator estimate \eqref{eq:eta} gives $\norm{K_j}\leq1$. Its definition uses only the known auxiliary query operators, the isometry $W$, and the degree projector.

Let $\ket{\psi_{j,x}}$ be the ideal joint state immediately before the $j$th auxiliary permutation query. Since each ideal auxiliary query increases degree by at most one, $$ \Pi_{\leq j-1}\ket{\psi_{j,x}} = \ket{\psi_{j,x}}. $$

Let $B_{j,x}$ denote the evolution of the sampled circuit strictly after the $j$th auxiliary query. Define $$ \ket{z_{j,x}} = B_{j,x}K_j\ket{\psi_{j,x}}. $$ Its norm is at most one because $K_j$ has norm at most one and the sampled evolution is unitary.

The map $W$ acts only on the external registers. All operations between auxiliary permutation queries act on the accessible registers and commute with $W$, as established above. In particular, this applies to the queries to $O_x$.

Compare successive circuits in which the first $j$ auxiliary queries are ideal and the remaining queries are sampled, with $W$ inserted between the two parts. The difference between the circuits for $j$ and $j-1$ is $$ B_{j,x}(WP_j-\widehat P_jW)\ket{\psi_{j,x}}. $$ Summing these differences gives the exact identity \begin{align} \ket{u_x}-\ket{v_x} &= \sum_{j=1}^q B_{j,x}(WP_j-\widehat P_jW)\ket{\psi_{j,x}} \nonumber\\ &= \eta\sum_{j=1}^q\ket{z_{j,x}}. \label{eq:state-residual} \end{align}

\paragraph{Step 2: realizing the cross terms within the query bound.}
Let $n_j$ be the number of input queries before the $j$th auxiliary permutation query. The construction in \cref{lem:aligned-preparation} realizes $K_j$ through an input-independent unitary dilation $U_{K_j}$, as shown in \cref{fig:residual-query-budget}. The preparations of $\ket{u_x}$ and $\ket{v_x}$ also use at most $T$ queries.

\begin{figure}[H]
\centering
\begin{tikzpicture}[
  queryblock/.style={draw,rounded corners=1pt,align=center,
    font=\small,text width=3.8cm,minimum height=1.05cm,inner sep=5pt}
]
\node[queryblock] (prefix) at (0,0)
  {Ideal prefix\\$n_j$ input queries};
\node[queryblock] (error) at (4.7,0)
  {Unitary dilation $U_{K_j}$\\$0$ input queries};
\node[queryblock] (suffix) at (9.4,0)
  {Sampled suffix\\$T-n_j$ input queries};
\draw[->,semithick] (prefix.east) -- (error.west);
\draw[->,semithick] (error.east) -- (suffix.west);
\end{tikzpicture}
\caption{A unitary preparation using $n_j+(T-n_j)=T$ input queries. Its final component with flag zero is $\ket0\ket{z_{j,x}}$. The flag is retained without measurement or postselection.}
\label{fig:residual-query-budget}
\end{figure}
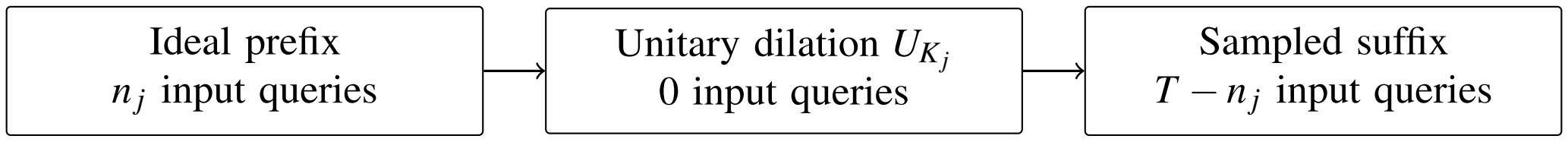

Let $M$ be the input-independent Hermitian operator defining the signed output expectation. It acts as the identity on the external registers and satisfies $\norm M\leq1$. Let $$ \ell_{\cA}^0(x) = \E_S a_{\cA}^0(S,x[S]) $$ be the expectation of the sampled circuit before the additional symmetrization below. The argument in \cref{lem:local-approximation} shows that $a_{\cA}^0$ depends only on $S$ and $x[S]$, with $|a_{\cA}^0|\leq1$ and $$ \bar h_{\cA}(x)=\bra{u_x}M\ket{u_x}, \qquad \ell_{\cA}^0(x)=\bra{v_x}M\ket{v_x}. $$

Since $M$ is Hermitian, \begin{align} \bar h_{\cA}(x)-\ell_{\cA}^0(x) &= \operatorname{Re}\bra{u_x-v_x}M\ket{u_x} + \operatorname{Re}\bra{v_x}M\ket{u_x-v_x} \nonumber\\ &= \eta\sum_{j=1}^q \left( \operatorname{Re}\bra{z_{j,x}}M\ket{u_x} + \operatorname{Re}\bra{v_x}M\ket{z_{j,x}} \right). \label{eq:expectation-residual} \end{align}

By \cref{lem:aligned-preparation}, the two preparations in each cross term share the same $T$ query positions. Thus each of the $2q$ real parts in \eqref{eq:expectation-residual} is the signed output expectation of an algorithm using at most $T$ queries.

Define $\cR^0(\cA)$ as follows. With probability $2q\eta/\lambda$, choose one of these $2q$ algorithms uniformly and run it; otherwise output a fair sign. This input-independent choice is valid because $2q\eta\leq\lambda$. Equation~\eqref{eq:expectation-residual} gives $$ h_{\cR^0(\cA)}(x) = \frac{\bar h_{\cA}(x)-\ell_{\cA}^0(x)}{\lambda}. $$

\paragraph{Step 3: symmetrizing the residual and the postprocessing function.}
Let $\cR(\cA)$ run $\cR^0(\cA)$ on $x^\tau$ for a uniformly random $\tau\in S_N$. Then $$ h_{\cR(\cA)}(x) = \E_{\tau\in S_N}h_{\cR^0(\cA)}(x^\tau). $$ This expectation is invariant under coordinate permutations. For each fixed $\tau$, a query to $x^\tau$ can be implemented by relabeling the address, applying $O_x$ once, and reversing the address relabeling. The relabeling operations are known and input-independent. Hence $\cR(\cA)$ still uses at most $T$ queries.

Let $U\subseteq[N]$ have size $s$. Let $z=(z_u)_{u\in U}$ be a collection of input values. Define $$ a_{\cA}(U,z) = \E_{\tau\in S_N} a_{\cA}^0\!\left(\tau^{-1}U, (z_{\tau(i)})_{i\in\tau^{-1}U} \right). $$ Thus $a_{\cA}(U,z)$ depends only on $U$ and $z$ and takes values in $[-1,1]$.

For each fixed $\tau$, a uniform $U$ is carried to a uniform $\tau^{-1}U$. Consequently, \begin{align*} \E_U a_{\cA}(U,x[U]) &= \E_\tau\E_S a_{\cA}^0\!\left( S,(x_{\tau(i)})_{i\in S} \right)\\ &= \E_\tau\ell_{\cA}^0(x^\tau). \end{align*} Also, $$ \bar h_{\cA}(x^\tau)=\bar h_{\cA}(x). $$ Averaging $$ \bar h_{\cA}(x^\tau) = \ell_{\cA}^0(x^\tau) + \lambda h_{\cR^0(\cA)}(x^\tau) $$ over $\tau$ proves \eqref{eq:residual}.
\end{proof}

Before iterating the residual construction, we check that the resulting algorithm still satisfies the hypotheses needed for another application.

\begin{remark}[Repeated application]
The residual algorithm is an ordinary query algorithm with at most $T$ queries to $O_x$. Its workspace may contain permutation registers, sample registers, and retained flags from its construction.

At the next application of \cref{lem:residual}, these registers are part of the algorithm's accessible workspace. The new random relabeling is represented by a separate external permutation register initialized independently. The operator estimate \eqref{eq:eta} allows arbitrary accessible workspaces and entanglement with them. It therefore applies at every recursion level.

\end{remark}

\subsection{One Sample for the Geometric Expansion}
\label{sec:geometric-expansion}

Iterating the residual identity gives a geometric expansion. We show that every term can be evaluated from the same sample, so retaining more terms does not require additional queries.

Let $\cA_0$ be the symmetrization of the original algorithm. For each integer $j\geq0$, let $$ \cA_{j+1}=\cR(\cA_j), \qquad h_j=h_{\cA_j}. $$ Every $\cA_j$ is symmetrized, so $h_j=\bar h_{\cA_j}$. Let $a_j$ be the postprocessing function supplied by \cref{lem:residual}. Then $$ h_j(x) = \E_Sa_j(S,x[S])+\lambda h_{j+1}(x). $$

Repeated substitution gives \begin{equation}\label{eq:geometric-residual} h_0(x) = \E_S\sum_{j=0}^{L-1} \lambda^j a_j(S,x[S]) + \lambda^Lh_L(x). \end{equation} The sample $S$ has the same distribution in every term. By linearity of expectation, we may therefore use a single sample $S$ and evaluate all terms from $x[S]$. Their evaluations need not be independent. In particular, the number of queries is $s$, not $Ls$.

Define \begin{equation}\label{eq:gL} g_L(S,x[S]) = (1-\lambda) \sum_{j=0}^{L-1}\lambda^j a_j(S,x[S]). \end{equation} The normalization ensures that this function can serve as a signed output expectation. Indeed, $$ |g_L(S,x[S])| \leq (1-\lambda)\sum_{j=0}^{L-1}\lambda^j = 1-\lambda^L \leq1. $$ Equation~\eqref{eq:geometric-residual} also gives \begin{equation}\label{eq:gL-expectation} \E_Sg_L(S,x[S]) = (1-\lambda) \bigl(h_0(x)-\lambda^Lh_L(x)\bigr). \end{equation}

Choosing the truncation depth and output precision gives the following guarantee.

\begin{proposition}[Bias-preserving simulation]
\label{prop:bias-string}
Let $T\geq1$ and $0<\beta\leq1/2$. Suppose a quantum algorithm using at most $T$ queries succeeds with probability at least $1/2+\beta$ on every promised input of a permutation-invariant function $f$. Then there is a nonadaptive randomized algorithm with success probability at least $1/2+3\beta/4$ using $O(T^2)$ queries.
\end{proposition}

The simulator uses a finite table determined by the geometric expansion. Set $L=\lceil\log_8(4/\beta)\rceil$ and $d=\lceil\log_2(32/\beta)\rceil$. For each possible sample $(S,z)$, fix $m_{S,z}\in\{0,\ldots,2^d\}$ so that $p(S,z)=m_{S,z}/2^d$ is a nearest grid point to $(1+g_L(S,z))/2$, with $g_L$ defined in \eqref{eq:gL}. The alphabet and the collection of samples are finite, so this table forms part of the simulator's input-independent description. A single run chooses $S$ uniformly, queries all $s$ coordinates in $S$, and outputs $1$ with probability $p(S,x[S])$.

\begin{proof}
We analyze a single run. Let $F(x)=2f(x)-1$. Symmetrization preserves the promise and the correct answer, so the initial symmetrized algorithm satisfies $$ F(x)h_0(x)\geq2\beta $$ on every promised input.

The choice of $L$ ensures $\lambda^L\leq\beta/4$. By \eqref{eq:gL-expectation} and $|h_L(x)|\leq1$, $$ F(x)\E_Sg_L(S,x[S]) \geq (1-\lambda)(2\beta-\lambda^L) \geq \frac{49}{32}\beta. $$

\paragraph{Finite output precision.}
The rounded signed expectation $\widetilde g_L=2p-1$ satisfies $$ |\widetilde g_L(S,z)-g_L(S,z)| \leq 2^{-d}\leq\frac{\beta}{32}. $$ Combining this rounding bound with the preceding estimate gives $$ F(x)\E_S\widetilde g_L(S,x[S]) \geq\frac{49}{32}\beta-\frac{\beta}{32} =\frac32\beta. $$ By \eqref{eq:signed-success}, the success probability is at least $1/2+3\beta/4$.

A single run uses $s$ nonadaptive queries. The parameters $L,d=O(\log(1/\beta))$ affect only its fixed decision rule and output randomness. For $T\geq1$, $$ s \leq (4K+1)T^2, $$ which proves the stated query bound.
\end{proof}

\subsection{Proof of Theorem~\ref{thm:string-main}}
\label{sec:error-restoration}

For $1/6\leq\eps<1/2$, taking the majority of three independent runs of the preceding simulator restores the original success guarantee. This increases the query cost by only a factor of three, independently of $\eps$.

\begin{stringrestatement}
For every finite alphabet $A$, every permutation-invariant function $f:\cD\subseteq A^N\to\{0,1\}$, and every $1/6\leq\eps<1/2$, there is a randomized algorithm that computes $f$ with error at most $\eps$ on every promised input using $O(Q_\eps(f)^2)$ queries.
\end{stringrestatement}

\Cref{alg:string-simulator} chooses all three samples before reading any input value.

\begin{algorithm}[H]
\caption{Classical simulation with the original error guarantee}
\label{alg:string-simulator}
\small
\begin{algorithmic}[1]
\Require Classical query access to $x\in\cD$; a $T$-query quantum algorithm $\cA$ for $f$ with error $\eps\in[1/6,1/2)$, where $T\geq1$.
\FixedData $s=\min\{N,\lceil K(2T)^2\rceil\}$, and the precision $d$ and table $(m_{S,z})$ from \cref{sec:geometric-expansion} for $\beta=1/2-\eps$.
\Ensure A bit $B$ satisfying $\Pr[B=f(x)]\geq1-\eps$.
\For{$r=1,2,3$}
  \State $S_r\gets$ an independent uniform element of $\binom{[N]}s$
\EndFor
\For{$r=1,2,3$}
  \State $z_r\gets (x_i)_{i\in S_r}$ \Comment{query and store the $s$ entries}
  \State $u_r\gets\operatorname{Uniform}\{0,\ldots,2^d-1\}$
  \State $B_r\gets\one[u_r<m_{S_r,z_r}]$
\EndFor
\State \Return $\operatorname{Majority}(B_1,B_2,B_3)$
\end{algorithmic}
\end{algorithm}

\begin{proof}
Fix $1/6\leq\eps<1/2$ and let $$ T=Q_\eps(f), \qquad \beta=\frac12-\eps. $$ Then $0<\beta\leq1/3$.

If $T=0$, then $f$ is constant on its promise and requires no randomized queries. Assume that $T\geq1$. By \cref{prop:bias-string}, each bit $B_r$ in \cref{alg:string-simulator} has success bias at least $3\beta/4$. On a fixed input with single-run bias $\alpha$, the majority of three independent runs has bias $m(\alpha)=3\alpha/2-2\alpha^3$. This function is nondecreasing on $[0,1/2]$. Since $0<\beta\leq1/3$, $$ m(\alpha)\geq m(3\beta/4) =\frac98\beta-\frac{27}{32}\beta^3 \geq\frac{33}{32}\beta \geq\beta. $$ Thus three independent runs restore success probability at least $1/2+\beta=1-\eps$.

The three runs use at most $$ 3s=O(T^2) $$ queries. Their sample sets and random choices can all be selected before observing the input values. The resulting simulator is nonadaptive, and the constant in its query bound is independent of $\eps$ throughout $1/6\leq\eps<1/2$.
\end{proof}

For completeness, consider $0<\eps<1/6$. Apply the bound just proved at error $1/3$ and take the majority of $O(\log(2/\eps))$ independent runs. A standard concentration bound reduces the error to $\eps$. Since $Q_{1/3}(f)\leq Q_\eps(f)$, this uses $$ O\!\left(Q_\eps(f)^2\log(2/\eps)\right) $$ queries. The resulting simulator remains nonadaptive.

For a power of two $n\geq2$, the permutation-invariant Deutsch--Jozsa problem $\mathrm{DJ}_n$ distinguishes constant strings in $\{0,1\}^n$ from strings of Hamming weight $n/2$. The exact quantum algorithm~\cite[Section~3]{CleveEtAl1998} and Bera's classical lower bound~\cite[Theorem~2]{Bera2015}, together with the standard sampling upper bound, give $$ Q_0(\mathrm{DJ}_n)=1, \qquad R_\eps(\mathrm{DJ}_n) =\Theta\!\left(\min\{n,\log(1/\eps)\}\right), \qquad 0<\eps\leq1/3. $$ Taking $n$ sufficiently large for each $\eps$ shows that the logarithmic factor in the preceding string simulation bound is necessary in general.

\section{Simulation of Graph and Hypergraph Properties}\label{sec:hypergraph-simulation}

We extend the string simulation to $k$-uniform hypergraph properties, with graph properties as the case $k=2$. Fix $k\geq2$ and a bound $T\geq1$ on the number of queries made by the quantum algorithm. We use the representation and query model from \cref{sec:graph-model}.

The simulator chooses a uniform set of $O_k(T^2)$ vertices and queries the induced hypergraph. The main difference from strings is that one hyperedge query requires relabeling $k$ vertices. All these relabelings use the same permutation, so the operator estimates and residual construction from the preceding sections apply to a single permutation register. We verify the compilation and the dependence on the sampled hypergraph, then use the residual identity to prove \cref{thm:hypergraph-main}.

\subsection{Compiling Hypergraph Queries}

For a hypergraph $H$ on $[N]$, the input oracle acts as $$ O_H\ket{v_1,\ldots,v_k,z} =\ket{v_1,\ldots,v_k,z\oplus H(v_1,\ldots,v_k)}. $$ The ordered representation assigns value zero whenever two vertex arguments coincide. We also extend the oracle to act as the identity if any argument is $\bot$. This extension uses one ordinary query as in \cref{lem:aligned-preparation}.

For a permutation $\pi\in S_N$, the relabeled input satisfies $$ H^\pi(v_1,\ldots,v_k)=H(\pi(v_1),\ldots,\pi(v_k)). $$

\begin{lemma}[Compilation of hypergraph queries]
\label{lem:hypergraph-compilation}
Let $\cA$ make at most $T$ hypergraph queries, and let $\pi$ be a uniform vertex permutation. Its execution on $H^\pi$ can be compiled into a circuit with exactly $T$ queries to $O_H$ and \begin{equation}\label{eq:hypergraph-q} q=2kT \end{equation} auxiliary queries to the same permutation oracle.
\end{lemma}

\begin{proof}
Pad the circuit to exactly $T$ queries as in \cref{sec:preliminaries}. For each query, allocate $k$ fresh auxiliary address registers initialized to $\ket\bot$ and retain them afterward.

To query $H^\pi$ on $(v_1,\ldots,v_k)$, compute $\pi(v_1),\ldots,\pi(v_k)$ into the fresh registers, apply $O_H$ once, and reverse the $k$ address computations. In the ideal execution, the auxiliary registers return to $\ket\bot$. Each computation and its reversal uses the permutation query from \cref{sec:compiled-relabeling}. Thus every query uses $2k$ auxiliary queries to the same permutation. The construction also applies to superpositions of vertex tuples. Only the $T$ queries to $O_H$ are charged.
\end{proof}

\subsection{Sampling Vertices}

We choose one vertex sample for all auxiliary permutation queries and show that the sampled circuit depends only on the induced hypergraph. Let $K=10^8$ and $q=2kT$. If $N\leq\lceil Kq^2\rceil$, query the entire hypergraph and compute the function exactly. This costs at most $$ \binom Nk\leq N^k\leq\lceil K(2kT)^2\rceil^k=O_k(T^{2k}). $$ For the rest of the construction, assume $N>\lceil Kq^2\rceil$ and set \begin{equation}\label{eq:vertex-sample} s=\lceil Kq^2\rceil. \end{equation} Choose $S$ uniformly among the $s$-element subsets of $[N]$. Write $H[S]$ for the induced hypergraph, namely the restriction of $H$ to $\binom Sk$. Reading $H[S]$ uses \begin{equation}\label{eq:induced-hypergraph-cost} \binom sk\leq s^k=O_k(T^{2k}) \end{equation} queries and determines every ordered query whose vertex arguments lie in $S$.

Use the isometry $W=J_sV_N$ and the sampled permutation queries from \cref{sec:string-sampling}, including their extensions to the ambient spaces. Let $\eta=\eta_{N,s,q}$. For each ideal auxiliary query $P$ with sampled counterpart $\widehat P$, the one-query estimate \eqref{eq:eta} gives \begin{equation}\label{eq:hypergraph-query-approximation} \norm{(\widehat PW-WP)\Pi_{\leq d}}\leq\eta, \qquad 0\leq d\leq q. \end{equation} The parameter choice in \eqref{eq:vertex-sample} satisfies the hypotheses of that estimate, and \eqref{eq:lambda-budget} yields $$ 2q\eta\leq\lambda,\qquad \lambda=\frac18. $$ After $d$ ideal auxiliary queries, the external permutation state has degree at most $d$.

\begin{lemma}[Dependence on the induced hypergraph]
\label{lem:hypergraph-locality}
Replace the auxiliary queries in the circuit of \cref{lem:hypergraph-compilation} by the sampled queries above. For every fixed sample $S$, the resulting circuit depends on the input only through $H[S]$. Its signed output expectation is a function $a_{\cA}^0(S,H[S])\in[-1,1]$, and the overall sampled expectation is $$ \ell_{\cA}^0(H)=\E_S a_{\cA}^0(S,H[S]). $$
\end{lemma}

\begin{proof}
Fix $S$. The address-support argument in \cref{lem:local-approximation} shows that each fresh auxiliary address register is supported on $\{\bot\}\cup S$ after its first sampled permutation query.

If an argument is $\bot$, the input oracle acts as the identity. Otherwise all arguments lie in $S$. Its response is then determined by $H[S]$, with value zero on repeated-vertex tuples. Hence each query can be answered from $H[S]$. Fresh address registers make the same argument valid at every query.

The initial sample label is uniform and all subsequent operations preserve it. The output measurement acts only on accessible registers, so the overall expectation is the uniform average over $S$.
\end{proof}

\subsection{A Residual Identity for Hypergraphs}

The preceding estimates allow us to apply the residual construction without increasing the query bound. For an algorithm $\cA$ querying hypergraphs, write $$ \bar h_{\cA}(H)=\E_{\pi\in S_N}h_{\cA}(H^\pi) $$ for its symmetrized signed expectation.

\begin{lemma}[Residual identity for hypergraphs]
\label{lem:hypergraph-residual}
Let $T\geq1$ and assume $N>\lceil K(2kT)^2\rceil$. For every algorithm $\cA$ making at most $T$ hypergraph queries, there are a function $a_{\cA}(S,H[S])\in[-1,1]$ and a symmetrized algorithm $\cR(\cA)$ making at most $T$ hypergraph queries such that \begin{equation}\label{eq:hypergraph-residual} \bar h_{\cA}(H) =\E_S a_{\cA}(S,H[S])+\lambda h_{\cR(\cA)}(H), \qquad \lambda=\frac18. \end{equation} Here $S$ is uniform among the $s$-element subsets of $[N]$, with $s$ as in \eqref{eq:vertex-sample}.
\end{lemma}

\begin{proof}
Apply the construction of \cref{lem:residual} to the compiled circuit from \cref{lem:hypergraph-compilation}, with $q=2kT$. The operators $K_j$ defined by \eqref{eq:Kj} are contractions by \eqref{eq:hypergraph-query-approximation}. The map $W$ commutes with every intervening operation, including $O_H$, and each ideal prefix and sampled suffix together use $T$ input queries. Thus Steps 1 and 2 of that proof apply. Since $2q\eta\leq\lambda$, they give an algorithm $\cR^0(\cA)$ using at most $T$ queries and satisfying $$ \bar h_{\cA}(H)=\ell_{\cA}^0(H)+\lambda h_{\cR^0(\cA)}(H). $$

To symmetrize, let $\cR(\cA)$ run $\cR^0(\cA)$ on $H^\tau$ for a uniform $\tau\in S_N$. A query to $H^\tau$ requires only a known relabeling of the $k$ vertex arguments and one query to $H$, so the query bound remains $T$.

For an $s$-element set $U$ and a hypergraph $Z$ on $U$, define $Z^\tau$ on $\tau^{-1}U$ by $Z^\tau(e)=Z(\tau(e))$. Set $$ a_{\cA}(U,Z) =\E_{\tau\in S_N}a_{\cA}^0(\tau^{-1}U,Z^\tau). $$ This function takes values in $[-1,1]$ and uses only the supplied hypergraph $Z$. For each fixed $\tau$, the set $\tau^{-1}U$ is a uniform $s$-element subset when $U$ is. Consequently, $$ \E_U a_{\cA}(U,H[U]) =\E_{\tau\in S_N}\ell_{\cA}^0(H^\tau). $$ Averaging the preceding residual identity over $\tau$ and using $\bar h_{\cA}(H^\tau)=\bar h_{\cA}(H)$ proves \eqref{eq:hypergraph-residual}.
\end{proof}

For a fixed original bound $T$, every residual algorithm uses at most $T$ queries, so the same values of $q$ and $s$ apply at every recursion level. Previous auxiliary registers become part of the accessible workspace, and the next application uses a fresh external permutation register, as in the string construction.

\begin{proposition}[Bias preservation for hypergraph properties]
\label{prop:bias-hypergraph}
Let $T\geq1$ and $0<\beta\leq1/2$. If a quantum algorithm using at most $T$ queries computes a $k$-uniform hypergraph property with success probability at least $1/2+\beta$ on every promised input, then a nonadaptive randomized algorithm succeeds with probability at least $1/2+3\beta/4$ using $O_k(T^{2k})$ queries.
\end{proposition}

\begin{proof}
The full-input case was handled above, so assume $N>\lceil Kq^2\rceil$. Symmetrization preserves the success guarantee. Iterate \cref{lem:hypergraph-residual} with the original bound $T$ fixed. The geometric expansion and probability rounding from \cref{prop:bias-string} apply with $x[S]$ replaced by $H[S]$, giving success probability at least $1/2+3\beta/4$ from one sampled induced hypergraph. Each residual uses the same uniform distribution on $s$-element vertex sets, so all terms can be evaluated from this one sample.

The only queries read $H[S]$, at cost at most $s^k=O_k(T^{2k})$. The set $S$ is chosen before observing the input, so the simulator is nonadaptive.
\end{proof}

\subsection{Proof of Theorem~\ref{thm:hypergraph-main}}
\label{sec:hypergraph-main-proof}

We now restore the original error guarantee using the same majority argument as for strings.

\begin{hypergraphrestatement}
For every fixed integer $k\geq2$, every $k$-uniform hypergraph property $f$, and every $1/6\leq\eps<1/2$, there is a randomized algorithm that computes $f$ with error at most $\eps$ on every promised input using $O_k(Q_\eps(f)^{2k})$ hyperedge-membership queries.
\end{hypergraphrestatement}

\begin{proof}
If $Q_\eps(f)=0$ and $\eps<1/2$, then $f$ is constant on its promise and requires no randomized queries. Otherwise, let $T=Q_\eps(f)\geq1$ and $\beta=1/2-\eps$. For $1/6\leq\eps<1/2$, we have $0<\beta\leq1/3$.

By \cref{prop:bias-hypergraph}, one randomized run has success bias at least $3\beta/4$ and uses $O_k(T^{2k})$ queries. The majority of three independent runs restores bias at least $\beta$, by the calculation in \cref{sec:error-restoration}. Thus $$ R_\eps(f)=O_k\!\left(Q_\eps(f)^{2k}\right). $$ All sample sets can be chosen in advance, so the simulator is nonadaptive. Taking $k=2$ gives the quartic bound for graph properties in the adjacency-matrix model.
\end{proof}

For the small-error extension, let $0<\eps<1/6$. Apply the theorem at error $1/3$ and take a majority vote over $O(\log(2/\eps))$ independent runs. Since $Q_{1/3}(f)\leq Q_\eps(f)$, this gives $$ R_\eps(f)=O_k\!\left(Q_\eps(f)^{2k}\log(2/\eps)\right). $$ The resulting simulator remains nonadaptive.

\section{Hypergraph Query Separations}\label{sec:lower-bound}

We now prove that the bound in \cref{thm:hypergraph-main} is nearly tight. Fix an integer $k\geq2$. We encode a Boolean promise problem into an uncolored connected simple $k$-uniform hypergraph. Structural recovery makes the source input independent of the vertex labeling, and the query reductions in \cref{prop:encoding} transfer its classical lower bound and quantum upper bound. Applying this encoding to Forrelation with increasing order will prove \cref{thm:lower-main}.

We use the representation and query model from \cref{sec:graph-model}. Throughout this section, $\widetilde O_k$ suppresses logarithmic factors in the relevant input size and constants depending only on $k$.

\subsection{Encoding Construction}

The encoding uses data vertices to store the source input and address vertices to identify each data vertex after relabeling.

Let $g:\cD_g\subseteq\{0,1\}^m\to\{0,1\}$ be nonconstant. Fix $b\geq b_0(k)$ and set $$ d=2^b-1,\qquad d_* = 2^{b-2}=\frac{d+1}{4},\qquad N=d+b. $$ The constant $b_0(k)\geq7$ will be chosen sufficiently large that $d\geq4k$ and the two vertex classes below have separated degree ranges. Assume $m\leq\binom{d_*}{k}$.

The vertex set is the disjoint union $$ V=X\mathbin{\cup}A,\qquad X=\{x_w:w\in\{0,1\}^b\setminus\{0^b\}\},\qquad A=\{a_1,\ldots,a_b\}. $$ The vertices in $A$ encode the binary addresses of the vertices in $X$. We store the source bits only on $k$-subsets of $$ X_*=\{x_w:w_1=w_2=1\}, $$ which has size $d_*$. This restriction separates the degrees of $X$ and $A$ while retaining $\Theta_k(d^k)$ possible source positions. Order $X_*$ by binary address, order its $k$-subsets lexicographically, and let $$ e:[m]\longrightarrow\binom{X_*}{k} $$ enumerate the first $m$ subsets.

On $A$, define the tree $\mathcal T$ with center $a_1$ and three paths $$ (a_1,a_2),\qquad (a_1,a_3,a_4),\qquad (a_1,a_5,a_6,\ldots,a_b). $$ Their lengths are $1$, $2$, and $b-4$, which are distinct because $b\geq7$. Every automorphism fixes the unique vertex of degree three, then each branch and every vertex on that branch. Thus $\mathcal T$ has only the identity automorphism and will identify the order of the address coordinates.

For $z\in\{0,1\}^m$ and $E\in\binom Vk$, define \begin{equation}\label{eq:hypergraph-encoding} H_z(E)= \begin{cases} z_j, & E=e(j),\\ \displaystyle\bigvee_{x_w\in Y}w_i, & E=\{a_i\}\cup Y,\quad Y\in\binom X{k-1},\\ \mathbf 1[\{a_i,a_j\}\in E(\mathcal T)], & E=\{a_i,a_j\}\cup C,\quad i<j,\quad C\in\binom X{k-2},\\ 0, & \text{otherwise}. \end{cases} \end{equation} Here $\bigvee$ denotes Boolean OR and $\mathbf 1[\cdot]$ is the indicator of the stated condition. The cases are disjoint, and every entry is either a known constant or one source bit.

For $k=2$, this is an ordinary graph: the pairs $e(j)\subseteq X_*$ store the source bits, the edge $\{a_i,x_w\}$ records $w_i$, and $A$ induces the rigid tree $\mathcal T$. For larger $k$, the second case records the OR of the corresponding address bits and the third case extends each tree edge with $k-2$ vertices of $X$. The sets $X$ and $A$ describe structural classes; no vertex colors or distinguished labels are supplied with the input.

\subsection{Structural Recovery}

We show that the hypergraph itself identifies both vertex classes and every binary address, so its isomorphism type determines the source input.

\begin{lemma}[Structural recovery]\label{lem:structural-recovery}
For sufficiently large $b_0(k)$, the following properties hold for every $b\geq b_0(k)$ and every $z$. The hypergraph $H_z$ is connected. There are a threshold $\tau$, determined only by $k$ and $b$, and a constant $\gamma_k>0$ such that $$ \deg(x)\leq\tau-\gamma_kd^{k-1}\quad(x\in X), \qquad \deg(a)\geq\tau+\gamma_kd^{k-1}\quad(a\in A). $$ The hypergraph uniquely identifies every $a_i$ and every address in $X$. Consequently, \begin{equation}\label{eq:isomorphism-z} H_z\cong H_{z'}\quad\Longleftrightarrow\quad z=z'. \end{equation}
\end{lemma}

\begin{proof}
After checking connectivity, we recover the vertex classes by degree, order the address coordinates using the rigid tree, and identify each data vertex from its address.

The hyperedges extending the edges of $\mathcal T$ connect all vertices in $A$. Every $x_w$ has a nonzero coordinate $w_i$, and hence belongs to a hyperedge containing $a_i$ and any $k-2$ other vertices of $X$. Thus $H_z$ is connected, independently of $z$.

To distinguish $X$ from $A$, we compare their degrees. Use the convention $\binom nt=0$ for $t<0$ or $t>n$. The three nonzero cases of \eqref{eq:hypergraph-encoding} give \begin{align} \deg(x_w) &\leq \binom{d_*-1}{k-1} +b\binom{d-1}{k-2} +(b-1)\binom{d-1}{k-3}, \label{eq:deg-X}\\ \deg(a_i) &= \binom d{k-1} -\binom{(d-1)/2}{k-1} +\deg_{\mathcal T}(a_i)\binom d{k-2}. \label{eq:deg-A} \end{align} In the second formula, exactly $(d-1)/2$ vertices of $X$ have their $i$-th address bit equal to zero. Define $$ c_X=\frac{4^{-(k-1)}}{(k-1)!},\qquad c_A=\frac{1-2^{-(k-1)}}{(k-1)!}. $$ Uniformly over $z,w,i$, the degree bounds imply $$ \frac{\deg(x_w)}{d^{k-1}}\leq c_X+O_k(b/d), \qquad \frac{\deg(a_i)}{d^{k-1}}=c_A+O_k(b/d). $$ For every $k\geq2$, we have $c_A>c_X$. Since $b/d\to0$, choose $b_0(k)$ so that the error terms are at most $(c_A-c_X)/4$ in absolute value. Then $$ \tau=\frac{c_X+c_A}{2}d^{k-1}, \qquad \gamma_k=\frac{c_A-c_X}{4} $$ have the stated properties.

Degrees therefore identify $X$ and $A$. Choose any $C_0\in\binom X{k-2}$. The values $H_z(C_0\cup\{a,a'\})$ for pairs of vertices in $A$ recover $\mathcal T$, whose rigidity identifies each $a_i$.

It remains to recover the address bits from the OR values. For $k=2$, set $C_i=\varnothing$ for every $i$. For $k\geq3$, choose any $Y_i\in\binom X{k-1}$ such that $$ H_z(\{a_i\}\cup Y_i)=0, $$ and let $C_i$ be any $(k-2)$-subset of $Y_i$. Such a $Y_i$ exists because $(d-1)/2\geq k-1$ vertices have their $i$-th bit equal to zero. Conversely, a zero answer certifies that every vertex in $Y_i$ has its $i$-th bit equal to zero. Writing $C_i=(c_{i,1},\ldots,c_{i,k-2})$ in any order, we obtain \begin{equation}\label{eq:recover-address} H_z(a_i,x_w,c_{i,1},\ldots,c_{i,k-2})=w_i \qquad\text{for every }x_w\in X. \end{equation} If $x_w\notin C_i$, this follows from the OR in \eqref{eq:hypergraph-encoding}. If $x_w\in C_i$, its $i$-th bit is zero and the ordered query has a repeated vertex, so its answer is also zero.

Every allowed choice of the sets $Y_i$ and $C_i$ therefore recovers the same binary addresses. An isomorphism preserves the classes, the ordered address vertices, and the zero answers used to choose these sets. By \eqref{eq:recover-address}, it sends each $x_w$ to the vertex with the same address. It consequently preserves each $e(j)$ and hence each $z_j$. This proves the forward implication in \eqref{eq:isomorphism-z}; the reverse implication is immediate.
\end{proof}

This structural recovery lets us define the invariant promise problem. Identify the canonical vertex set with $[N]$ and use the relabeling convention $H^\pi(E)=H(\pi(E))$ for $\pi\in S_N$. Define $$ \cD_{F_g}=\{H_z^\pi:z\in\cD_g,\ \pi\in S_N\}, \qquad F_g(H_z^\pi)=g(z). $$ We suppress the fixed parameter $b$ in the notation $F_g$. By \cref{lem:structural-recovery}, different representations of the same hypergraph give the same source input. Thus $F_g$ is well defined, its promise is closed under vertex permutations, and its value is unchanged by them.

\subsection{Query Reductions}
\label{sec:encoding-reductions}

The classical reduction uses only the canonical labeling, where each hyperedge query can be answered using at most one source query. For the quantum reduction, we must also handle arbitrary vertex relabelings: we first recover the address vertices and then locate each data vertex coherently when it is needed.

\begin{lemma}[Classical reduction]\label{lem:classical-reduction}
For every $b\geq b_0(k)$ satisfying $m\leq\binom{2^{b-2}}k$ and every $0<\eps<1/2$, $$ R_\eps(F_g)\geq R_\eps(g). $$
\end{lemma}

\begin{proof}
Simulate an algorithm for $F_g$ on the canonical labeling of $H_z$. Every entry in \eqref{eq:hypergraph-encoding} is a known constant or one source bit, and repeated-vertex tuples have the known value zero. Each hypergraph query therefore costs at most one query to $z$, with no change in the error probability.
\end{proof}

For the quantum reductions, let $H$ be an arbitrary relabeling of $H_z$. Write $X(H)$ and $A(H)$ for its structural classes, $\alpha_i$ for the input label corresponding to $a_i$, and $\xi_w$ for the input label of the vertex with address $w$. These labels are uniquely determined by \cref{lem:structural-recovery} but are initially unknown to the algorithm.

\begin{lemma}[Preprocessing]\label{lem:encoding-preprocessing}
There is a quantum preprocessing procedure with worst-case query cost $$ O_k(\sqrt{Nb}\log N+b^2). $$ It either aborts or returns fixed sampled sets defining a deterministic classifier $\widehat c$, candidate labels for $\alpha_1,\ldots,\alpha_b$, and sets $C_1,\ldots,C_b$ of size $k-2$. With probability at least $0.99$, the classifier correctly identifies both classes on every vertex, all address labels are correct, and $$ C_i\subseteq X(H),\qquad \text{every vertex in }C_i\text{ has its }i\text{-th address bit equal to zero}. $$ A reversible evaluation and uncomputation of $\widehat c(v)$ uses $O_k(\log N)$ hypergraph queries.
\end{lemma}

\begin{proof}
We build a fixed classifier, recover the ordered address vertices, and select padding sets for the coherent source queries in \cref{lem:source-query-simulation}.

\paragraph{Classifying vertices.}
Independently sample $L_0=\lceil K_k\log N\rceil$ uniform sets $D_1,\ldots,D_{L_0}\in\binom V{k-1}$ and keep them fixed. For each vertex $v$, let $Y_t(v)=H(v,D_t)$ and define $$ \widehat c(v)= \begin{cases} A,& L_0^{-1}\sum_{t=1}^{L_0}Y_t(v)\geq\theta,\\ X,& \text{otherwise}, \end{cases} \qquad \theta=\frac{\tau}{\binom N{k-1}}. $$ The classifier is evaluated reversibly on demand. For each fixed $v$, the variables $Y_t(v)$ are independent and satisfy $$ \mathbb E[Y_t(v)] =p_v:=\frac{\deg_H(v)}{\binom N{k-1}}. $$ By \cref{lem:structural-recovery} and $N=\Theta(d)$, every true mean is at distance at least a constant $\Delta_k>0$ from $\theta$. Hoeffding's inequality \cite[Theorem~2]{Hoeffding1963} and a union bound show that the event $\mathcal E$ on which every vertex is classified correctly satisfies $$ \Pr[\mathcal E^c]\leq 2N\exp(-2L_0\Delta_k^2)\leq\frac1{200}. $$ Choose $K_k$ so that $2L_0\Delta_k^2\geq\ln(400N)$.

The fixed samples define a deterministic classifier $\widehat c$ whose reversible evaluation and uncomputation use $O_k(\log N)$ queries. On $\mathcal E$, the same classifier is correct on all vertices, including when evaluated in superposition.

\paragraph{Finding and ordering the address vertices.}
Start with an empty list $B$. For $j=b,b-1,\ldots,1$, apply exact amplitude amplification to the predicate $\widehat c(v)=A$ and $v\notin B$ with prescribed solution count $j$. After each search, measure a vertex label and append it to $B$. Abort unless the $b$ labels are distinct and all classified as $A$. On $\mathcal E$, each search has exactly $j$ solutions and succeeds using $O(\sqrt{N/j})$ predicate evaluations \cite[Theorem~4]{BrassardEtAl2002}. Thus the recovered labels are exactly $A(H)$, and the total search cost is $$ O_k(\log N)\sum_{j=1}^bO(\sqrt{N/j}) =O_k(\sqrt{Nb}\log N). $$

Let $X'=V\setminus B$ be the candidate complement and choose any $C_0\in\binom{X'}{k-2}$. Query $H(u,v,C_0)$ for every pair $\{u,v\}\in\binom B2$ to obtain a graph $G$ on $B$. Abort unless $G\cong\mathcal T$. Otherwise, use the unique isomorphism to order the labels as $\alpha_1,\ldots,\alpha_b$. On $\mathcal E$, we have $X'=X(H)$, so these queries reproduce $\mathcal T$ and give the correct ordering. The verification and tree reconstruction use $O_k(b\log N+b^2)$ queries.

\paragraph{Choosing the sets $C_i$.}
For $k=2$, take $C_i=\varnothing$ for every $i$. For $k\geq3$, set $J=\lceil4^{k-1}\ln(300b)\rceil$. For each $i$, sample independent uniform sets $Y\in\binom{X'}{k-1}$ and query $H(\{\alpha_i\}\cup Y)$ until the first zero answer or $J$ trials have been made. Abort if all $J$ answers are one. Otherwise, choose $C_i$ as any $(k-2)$-subset of the first zero-answer set. This stage uses at most $bJ=O_k(b\log b)$ queries.

Conditional on $\mathcal E$, the complement equals $X(H)$. For each coordinate, exactly $q=(d-1)/2$ of these vertices have address bit zero. The probability of a zero answer in each trial is therefore $$ \rho=\frac{\binom q{k-1}}{\binom d{k-1}} =\prod_{j=0}^{k-2}\frac{q-j}{d-j} \geq4^{-(k-1)}, $$ where the last inequality uses $d\geq4k$. The probability of finding no such set for some coordinate is at most $$ b(1-\rho)^J\leq b\exp(-\rho J)\leq\frac1{300}. $$ Every zero answer certifies that all vertices in the selected $C_i$ have their $i$-th bit equal to zero, as in \eqref{eq:recover-address}. If no stage aborts, return the fixed samples, ordered address labels, and sets $C_i$. The probability of returning correct preprocessing data is at least $$ 1-\frac1{200}-\frac1{300}>0.99. $$

All searches use predetermined amplification counts and all sampling loops have fixed budgets. Its query bound therefore also holds on $\mathcal E^c$. Since $b=\Theta(\log N)$ and $b\log b=O(b^2)$, the total worst-case cost is $O_k(\sqrt{Nb}\log N+b^2)$.
\end{proof}

The preprocessing data allow us to recognize a vertex from its address. The next lemma turns this test into a coherent implementation of the source oracle.

\begin{lemma}[Coherent source query]\label{lem:source-query-simulation}
Suppose the preprocessing data satisfy the correctness conditions of \cref{lem:encoding-preprocessing}. Then the source oracle $$ O_z\ket{j,t}=\ket{j,t\oplus z_j} $$ can be implemented exactly using $O_k(\sqrt N\log N)$ hypergraph queries, with all auxiliary registers returned to their initial states. The circuit has the same query bound for any candidate data returned by preprocessing.
\end{lemma}

\begin{proof}
We implement a source query by coherently locating its $k$ data vertices, querying the resulting hyperedge, and reversing the location procedures. We first build the unitary for locating one vertex, then show that the complete computation restores every auxiliary register.

\paragraph{Locating a data vertex.}
For a nonzero binary address $w$, let $P_w(v)=1$ precisely when $\widehat c(v)=X$ and $$ H(\alpha_i,v,c_{i,1},\ldots,c_{i,k-2})=w_i \qquad\text{for every }i\in[b], $$ where each $C_i=(c_{i,1},\ldots,c_{i,k-2})$ is ordered arbitrarily. Compute the classifier and the $b$ query values, compare them with $w$, and uncompute the intermediate data. This reversible predicate uses $O_k(\log N)$ queries. By \eqref{eq:recover-address}, it has exactly one solution $\xi_w$.

Starting from a uniform superposition of the $N$ input labels, exact amplitude amplification \cite[Theorem~4]{BrassardEtAl2002} gives a unitary \begin{equation}\label{eq:address-unitary} U:\ket w\ket0\ket0_{\mathrm{work}} \longmapsto e^{i\theta_w}\ket w\ket{\xi_w}\ket0_{\mathrm{work}} \end{equation} using $O_k(\sqrt N\log N)$ queries. The marked fraction is always $1/N$, so the amplification counts and phase corrections are independent of $w$. The same circuit therefore works coherently on superpositions of addresses. Its inverse has the same query cost.

\paragraph{Implementing the source oracle.}
Given $j$, compute the $k$ binary addresses in $e(j)$ and apply $U$ to each. Query the resulting hyperedge and apply $U^\dagger$ to remove the recovered labels. Finally, uncompute the binary addresses. The hyperedge answer is $z_j$. The query changes only the answer bit, so each address-dependent phase in \eqref{eq:address-unitary} cancels under $U^\dagger$ and every auxiliary register is restored. This implements $O_z$ exactly, using $O_k(\sqrt N\log N)$ queries because $k$ is fixed.

For incorrect candidate data, the same fixed sequence of unitary operations is still defined and has the same query bound.
\end{proof}

\subsection{Encoding Proposition}

\begin{proposition}[Uncolored hypergraph encoding]
\label{prop:encoding}
Let $g:\cD_g\subseteq\{0,1\}^m\to\{0,1\}$ be nonconstant. For every $b\geq b_0(k)$ satisfying $m\leq\binom{2^{b-2}}k$, set $d=2^b-1$ and $N=d+b$. The function $F_g$ defined above is permutation-invariant on a promise of uncolored connected simple $k$-uniform hypergraphs with $N$ vertices and satisfies \begin{equation}\label{eq:encoding-classical} R_\eps(F_g)\geq R_\eps(g),\qquad0<\eps<1/2, \end{equation} and \begin{equation}\label{eq:encoding-quantum-N} Q_{1/3}(F_g) =O_k\!\left(\sqrt{Nb}\log N+b^2+ \sqrt N\log N\,Q_{1/3}(g)\right). \end{equation} Choosing the smallest admissible $b$ gives $N=\Theta_k(m^{1/k})$ and \begin{equation}\label{eq:encoding-quantum} Q_{1/3}(F_g) =\widetilde O_k\!\left(m^{1/(2k)}Q_{1/3}(g)\right). \end{equation}
\end{proposition}

\begin{proof}
The promise and invariance properties follow from \cref{lem:structural-recovery}, and \cref{lem:classical-reduction} gives the classical lower bound. We obtain the quantum bound by combining preprocessing with the coherent source oracle, then choose the smallest admissible $b$ to express the cost in terms of $m$.

For the quantum bound, run the preprocessing procedure of \cref{lem:encoding-preprocessing}. If it aborts, output an arbitrary bit. Otherwise, amplify a quantum algorithm for $g$ to error at most $1/12$ with a constant number of repetitions and replace each source query using \cref{lem:source-query-simulation}. With probability at least $0.99$, the preprocessing data are correct and every source query is simulated exactly. The total error is at most $0.01+1/12<1/3$. Combining the query bounds in \cref{lem:encoding-preprocessing,lem:source-query-simulation} proves \eqref{eq:encoding-quantum-N}.

For the final assertion, choose the smallest $b\geq b_0(k)$ with $\binom{2^{b-2}}k\geq m$. Since $\binom{2^{b-2}}k=\Theta_k(d^k)$, minimality gives $$ d=\Theta_k(m^{1/k}),\qquad N=d+b=\Theta_k(m^{1/k}), $$ with the lower cutoff $b_0(k)$ absorbed into the constants. Since $b=\Theta(\log N)$, substitution in \eqref{eq:encoding-quantum-N} gives \eqref{eq:encoding-quantum}.
\end{proof}

\subsection{Forrelation with Increasing Order}

To obtain the desired hypergraph separation, we need source problems with nearly linear randomized query complexity and polylogarithmic quantum query complexity. We use Forrelation with increasing order, so the dependence on that order must remain explicit.

\begin{lemma}[Forrelation separation]
\label{lem:forrelation-source}
For every integer $\ell\geq2$ and every power of two $L\geq2^{20\ell}$, there is an explicit nonconstant partial Boolean function $g_{\ell,L}$ on $m=\ell L$ bits such that \begin{equation}\label{eq:forrelation-Q} Q_{1/3}(g_{\ell,L})=O(\ell2^{10\ell}), \end{equation} and, for every $0<\beta<1/2$, \begin{equation}\label{eq:forrelation-R} R_{1/2-\beta}(g_{\ell,L}) = \Omega\!\left( \frac{\beta^2}{\ell^{29}} \left(\frac{L}{\log(\ell L)}\right)^{1-1/\ell} \right). \end{equation} The implicit constants are independent of $\ell$, $L$, and $\beta$.
\end{lemma}

\begin{proof}
Let $\delta=2^{-5\ell}$ and let $g_{\ell,L}$ be the $(\delta,\ell)$-Forrelation problem of Bansal and Sinha~\cite{BansalSinha2021}, with its $\ell L$ sign coordinates encoded as bits. The quantum bound \eqref{eq:forrelation-Q} follows from \cite[Corollary~1.4]{BansalSinha2021}, and \cite[Theorem~1.3]{BansalSinha2021} with $\eta=2\beta$ gives \eqref{eq:forrelation-R}.

The assumption $L\geq2^{20\ell}$ ensures $$ \frac4{\delta^2L}\leq2^{2-10\ell}\leq\delta. $$ Thus the uniform-input exceptional probability in \cite[Theorem~3.1]{BansalSinha2021} is at most $\delta$. The remaining parameter dependence is explicit in these bounds, so the constants are uniform in $\ell$, $L$, and $\beta$.
\end{proof}

\subsection{Proof of Theorem~\ref{thm:lower-main}}
\label{sec:lower-main-proof}

We apply \cref{prop:encoding} to the source problems in \cref{lem:forrelation-source}.

\begin{lowerrestatement}
For every fixed integer $k\geq2$, there is an explicit infinite family $\{F_N\}$ of $k$-uniform hypergraph properties with promises consisting of connected, uncolored, simple hypergraphs on $N$ vertices such that $$ R_{1/3}(F_N)\geq N^{k-o(1)}, \qquad Q_{1/3}(F_N)\leq N^{1/2+o(1)}. $$
\end{lowerrestatement}

\begin{proof}
We choose the source input size to fill a constant fraction of the available data hyperedges and let the Forrelation order grow slowly. The source then has polylogarithmic quantum query complexity and nearly linear randomized query complexity.

For each sufficiently large integer $b$, let $d=2^b-1$, $d_*=2^{b-2}$, and $$ \ell= \lfloor\log_2\log_2d\rfloor. $$ Let $L$ be the largest power of two satisfying $$ \ell L\leq\binom{d_*}{k}. $$ Maximality implies $$ \frac12\binom{d_*}{k}<\ell L\leq\binom{d_*}{k}. $$ Apply \cref{prop:encoding} to $g_{\ell,L}$ with this $b$ and $m=\ell L$. Let $F_N$ denote the resulting function on $$ N=d+b $$ vertices. The parameters satisfy $$ \begin{gathered} d_*=\Theta(d),\qquad d=\Theta(N),\qquad b=\Theta(\log N),\qquad m=\Theta_k(d^k),\\ \ell=\Theta(\log\log d),\qquad L=\Theta_k(d^k/\ell). \end{gathered} $$ This defines an explicit infinite family along the increasing sequence of vertex counts obtained from $b$.

For all sufficiently large $d$, $$ 2^{20\ell}\leq(\log_2d)^{20}. $$ Since $L=\Theta_k(d^k/\ell)$, it follows that $L\geq2^{20\ell}$. Thus \cref{lem:forrelation-source} applies.

For the quantum upper bound, $$ Q_{1/3}(g_{\ell,L}) = O(\ell2^{10\ell}) = O\!\left(\ell(\log d)^{10}\right). $$ Substituting these bounds into \eqref{eq:encoding-quantum-N} gives $$ Q_{1/3}(F_N) \leq C_k\sqrt N\,(\log N)^{12} = N^{1/2+o(1)} $$ for a constant $C_k$ and all sufficiently large $N$ in the family.

Since $L=\Theta_k(d^k/\ell)$ and $\log(\ell L)=\Theta_k(\log d)$, \eqref{eq:encoding-classical} and \eqref{eq:forrelation-R} imply $$ R_{1/2-\beta}(F_N) \geq c_k\frac{\beta^2}{\ell^{29}} \left( \frac{d^k}{\ell\log d} \right)^{1-1/\ell} $$ for some $c_k>0$ and every $0<\beta<1/2$.

Since $\ell\to\infty$ and $\ell=O(\log\log d)$, $$ \frac1{\ell^{29}} \left(\frac{d^k}{\ell\log d}\right)^{1-1/\ell} =d^{k-o(1)}. $$ Using $N=\Theta_k(d)$ and absorbing fixed constants into the $o(1)$ term gives $$ R_{1/2-\beta}(F_N)\geq\beta^2N^{k-o(1)}, $$ with an $o(1)$ term independent of $\beta$.

Taking $\beta=1/6$ gives the claimed bounded-error lower bound.
\end{proof}

The same family also satisfies $$ R_{1/2-\beta_N}(F_N)\geq N^{k-o(1)}, \qquad Q_{1/2-\beta_N}(F_N)\leq N^{1/2+o(1)} $$ for every sequence with $0<\beta_N\leq1/6$ and $\beta_N=N^{-o(1)}$. The randomized lower bound follows from $\beta_N^2=N^{-o(1)}$ in the preceding proof. The quantum upper bound follows from $Q_{1/2-\beta_N}(F_N)\leq Q_{1/3}(F_N)$.

\section*{Acknowledgments}
\phantomsection
\addcontentsline{toc}{section}{Acknowledgments}
We thank Penghui Yao for helpful discussions. Yunqi Huang  and  Shenggen Zheng were supported by the Guangdong Provincial Quantum Science Strategic Initiative (Grant No.~GDZX2503001) and the Shenzhen Science and Technology Program (Grant No.~ZDYJ20251211120859002). Lvzhou Li was supported by the National Key Research and Development Program of China (Grant No.~2024YFB4504004), the National Natural Science Foundation of China (Grant Nos.~92465202 and 62272492), the Guangdong Provincial Quantum Science Strategic Initiative (Grant No.~GDZX2303007), and the Guangzhou Science and Technology Program (Grant No.~2024A04J4892). Zekun Ye was supported by start-up research funding from Fuzhou University (Grant No.~XRC-26004).

\section*{AI Methodology}
We used OpenAI ChatGPT (Codex) from August to November 2026 to assist with developing and organizing the proofs, checking intermediate calculations, and reviewing and revising the manuscript. This assistance included the sampling argument in \cref{sec:string-sampling}, the residual expansion in \cref{sec:small-bias}, the hypergraph simulation in \cref{sec:hypergraph-simulation}, and the hypergraph encoding in \cref{sec:lower-bound}. The tool also assisted with numerical consistency checks of finite-dimensional instances of the restriction identity in \cref{lem:exact-norm}. The authors take full responsibility for the correctness, originality, and integrity of the paper, including its proofs, references, and attribution.

\bibliographystyle{alpha}
\begingroup
\raggedright
\bibliography{references}
\endgroup

\end{document}